\documentclass[article,showpacsreprint,groupedaddress,12pt,superscriptaddress,preprintnumbers,floatfix,a4paper,nofootinbib,amsmath,amssymb]{revtex4-2}
\usepackage[english]{babel}
\usepackage[letterpaper,top=2cm,bottom=2cm,left=2cm,right=2cm,marginparwidth=1.75cm]{geometry}
\usepackage{amssymb, amsmath, xcolor, graphicx}
\usepackage{pgf}
\usepackage{hyperref}
\hypersetup{%
pdftitle = {title},
pdfsubject = {},
pdfkeywords = {},
colorlinks = {true},
filecolor = {black},
linkcolor = {link_blue},
menucolor = {black},
citecolor = {link_blue},
urlcolor = {link_blue},
}{}

\usepackage{multirow}

\newcommand\bef{\begin{figure}}
\newcommand\eef[1]{\label{fg:#1}\end{figure}}
\newcommand\beq{\begin{equation}}
\newcommand\eeq[1]{\label{#1}\end{equation}}
\newcommand\beqa{\begin{eqnarray}}

\newcommand\bet{\begin{table}}
\newcommand\eet[1]{\label{tb:#1}\end{table}}

\definecolor{link_blue}{RGB}{51,102,204}

\begin{document}
\title{Quark-Mass Dependence of Light-Nuclei Masses from Lattice QCD\\
and Trace-Anomaly Contributions to Nuclear Bindings}
\author{Debsubhra Chakraborty}
%{~\orcidlink{0000-0001-5815-4182}}
\email{debsubhra.chakraborty@tifr.res.in}
\affiliation{Department of Theoretical Physics, Tata Institute of Fundamental Research, \\ Homi Bhabha Road, Mumbai 400005, India }

\author{Noah Chavez}
%\email{noah.chavez@stonybrook.edu}
\affiliation{Department of Physics and Astronomy, Stony Brook University, Stony Brook, New York 11794, USA}

\author{Xiang Gao}
\email{xgao@bnl.gov}
\affiliation{Physics Department, Brookhaven National Laboratory, Upton, New York 11973, USA }

\author{Nilmani Mathur}
%~\orcidlink{0000-0003-2422-7317}}
%\email{nilmani@theory.tifr.res.in}
\affiliation{Department of Theoretical Physics, Tata Institute of Fundamental Research, \\ Homi Bhabha Road, Mumbai 400005, India }

\author{Swagato Mukherjee}
%\email{swagato@bnl.gov}
\affiliation{Physics Department, Brookhaven National Laboratory, Upton, New York 11973, USA }

\preprint{TIFR/TH/26-11}

\begin{abstract}
\centerline{}

We present lattice QCD calculations of the masses of the deuteron, dineutron, Helium-3 and Helium-4 with physical sea quarks and valence quark masses corresponding to pion masses between 140 and 700 MeV. At the physical point, the lowest finite-volume two-nucleon energy levels exhibit the qualitative pattern of a bound deuteron and an unbound dineutron within uncertainties,while at heavier quark masses they indicate the presence of deeply bound states. Compared with expectations from low-energy effective field theories, the observed mass dependence of the binding energies provides first-principles constraints on the quark-mass dependence of two- and three-nucleon interactions. From the quark-mass variation of the nuclear energies, we determine nuclear sigma terms and quantify the response of light-nuclear masses to changes in the light-quark mass. Using the QCD trace anomaly relation, we decompose the nuclear binding energy into quark-mass and gluonic contributions around the deuteron mass scale of $\mu=2$~GeV. We find that the quark-mass contribution to the binding energy is small and approximately additive in nucleon number within current precision, whereas the gluonic component provides the dominant contribution and show milder increases with mass number.

\end{abstract}
\maketitle
\section{\label{sec:introduction}Introduction}
The origin of visible mass in the universe lies in the dynamics of quantum chromodynamics (QCD). While hadron masses arise predominantly from gluonic interactions rather than the bare quark masses, the trace of the QCD energy–momentum tensor provides an operator relation that separates quark-mass and gluonic contributions to the mass of a hadronic state~\cite{PhysRevLett.74.1071, Hatta:2018sqd,Tanaka:2018nae}. This decomposition has been studied extensively for individual hadrons both theoretically and in lattice QCD calculations~\cite{Bollweg:2026nnr,Tanaka:2025hhv,Chen:2025iul,Tanaka:2022wzy,Tanaka:2022wrr,Sun:2020ksc,PhysRevLett.121.212001, PhysRevD.109.094504}. Atomic nuclei are bound states of nucleons whose masses and interactions ultimately arise from QCD. Although nuclear binding energies are small compared with nucleon masses, they determine nuclear structure and the stability of matter. Understanding how nuclear binding emerges from QCD dynamics, and how quark- and gluon-operator contributions enter the binding energy, remains a central problem in nuclear physics. Addressing this question requires first-principles calculations of nuclear masses and their dependence on the light-quark masses.

Lattice QCD provides a framework for computing hadronic observables directly from the QCD Lagrangian. Significant progress has been made in lattice calculations of multi-nucleon systems~ \cite{Beane:2010em, Beane:2015yha, NPLQCD2013,NPLQCD2015,CalLat2017,Detmold:2015daa,PhysRevD.87.114512,Savage:2016kon, Wagman:2017tmp,Detmold:2019ghl, 
Davoudi:2020ngi,Detmold:2020snb,Parreno:2021ovq,NPLQCD2021,Davoudi:2024ukx,
Ishii:2006ec,Nemura:2008sp,Yamazaki_2010,Aoki:2009ji,Doi:2011gq,Aoki:2012tk,Gongyo:2017fjb, Francis:2018qch,
Drischler:2019xuo, Junnarkar:2019equ,green_weakly_2021,CalLat2021,Junnarkar:2022yak,Mathur:2022ovu, PhysRevD.110.114505}. However, most studies have been performed at pion masses heavier than the physical value, and the interpretation of multi-nucleon energy levels remains the subject of ongoing discussion \cite{PhysRevD.96.034521, BaSc2025, Detmold:2026vjx}. Calculations with physical sea quarks are therefore essential for establishing a direct connection between lattice QCD results and nuclear phenomena in nature.

The quark-mass dependence of nuclear energies provides information beyond binding energies at a single point in parameter space of strong interaction. Through the Feynman–Hellmann relation, the derivative of the nuclear mass with respect to the light-quark mass yields the nuclear sigma term, which quantifies the quark-mass contribution to the nuclear mass~\cite{Ohki:2008ff,Bali:2012qs,Bouchard:2016heu}. Determining this response provides direct information about the role of explicit chiral symmetry breaking in nuclear structure. The resulting mass dependence can also be compared with expectations from low-energy effective field theories (EFTs), thereby constraining the quark-mass dependence of two- and three-nucleon interactions~\cite{PhysRevC.102.044003,PhysRevD.103.074503,PhysRevD.105.074508,PhysRevD.108.034509,PhysRevD.109.114515}.

Furthermore, the trace of the QCD energy–momentum tensor relates the mass of a hadronic state to matrix elements of quark-mass and gluonic operators~\cite{Hatta:2018sqd,Tanaka:2018nae,Tanaka:2022wzy,Tanaka:2025hhv}. Applied to nuclei, this relation allows the nuclear mass, and hence the nuclear binding energy, to be decomposed into contributions associated with the quark-mass term and the gluonic term arising from the trace anomaly. By combining lattice determinations of nuclear sigma terms with physical nuclear masses, the gluonic component of nuclear binding can be inferred in a given renormalization scheme and scale. This provides a quantitative probe of the QCD operator structure underlying nuclear binding.

In this article, we present lattice QCD calculations of the masses of the deuteron ($^2\mathrm{H}$), dineutron (\textit{nn}), Helium-3 ($^3\mathrm{He}$), and Helium-4 ($^4\mathrm{He}$) using ensembles with physical sea quarks and valence quark masses corresponding to pion masses between 140 and 700 MeV. These calculations allow us to study the quark-mass dependence of the lowest finite-volume energies and to determine nuclear sigma terms from the response of nuclear energies to variations of the light-quark mass. The resulting mass dependence enables a decomposition of the nuclear binding energy into quark-mass and gluonic components through the QCD trace anomaly relation, while providing constraints on the quark-mass dependence of low-energy nuclear interactions.

\section{Simulation Details}
\subsection{\label{subsec:LatticeSetup}Lattice setup}
We use a $2+1$-flavor HISQ gauge ensemble generated by the HotQCD Collaboration~\cite{Follana:2006rc,HotQCD:2014kol, Bazavov:2019www}. The light sea-quark masses correspond to 140 MeV pion mass and physical strange quark mass. The lattice spacing is $a \approx 0.076$~fm on a $64^4$ lattice, corresponding to a physical volume of approximately $(5\,\mathrm{fm})^3$ (further details are in Appendix \ref{numerical_setup}). For the valence quarks we employ a tree-level tadpole-improved Wilson–clover action with one-step HYP-smeared gauge links entering the Dirac and clover operators~\cite{Hasenfratz:2001hp,Izubuchi:2019lyk,Gao:2020ito,Gao:2021hvs,Gao:2021xsm,Gao:2021dbh,Gao:2022vyh,Gao:2022iex,Gao:2022uhg,Gao:2023lny,Gao:2023ktu,Ding:2024lfj,Cloet:2024vbv,Ding:2024saz,Gao:2024fbh,Bollweg:2025iol,Gao:2025inf,Gao:2026hix}.

\subsection{\label{subsec:EnergyExtraction} Energy extractions}
We determine the finite-volume lowest-lying energies $E_A$, for light nuclei corresponding to the interpolating operators (see Appendix \ref{sec:op_const}), using both standard one- and two-exponential fits as well as the recently developed lattice transfer-matrix generalized eigenvalue problem (TGEVP) procedure of Refs.~\cite{Chakraborty:2024scw} (see Appendix \ref{subsec:TGEVP} for details).
%formalism in a Krylov subspace \cite{Lanczos,BlockLanczos,Ostmeyer:2024qgu,Chakraborty:2024scw} 
For exponential fits, the fit windows are varied within the effective-mass plateau region to test stability and estimate fit-window uncertainties. For the TGEVP procedure, we remove spurious eigenvalues robustly by determining threshold levels based on adaptive kernel density estimation (KDE)~\cite{Silverman1998,10.1214/aos/1176342997,Hastie2009} of the bootstrapped eigenvalue distributions. We ensure two methods give consistent results. Using the extracted the single-nucleon mass $m_N$ and $E_A$, we determine the energy splittings, $\Delta E_A = E_A - A\,m_N$, for the finite-volume lowest-lying states.
All analyses are performed within a bootstrap framework to account for statistical correlations. For details of the energy extractions see Appendix \ref{sec:DetailsEnergyExtraction}. The finite-volume energy splitting $\Delta E_A$ is related to the nuclear binding energy once a finite-volume amplitude analysis of the lattice energy levels is performed.

\begin{figure}[hbt!]
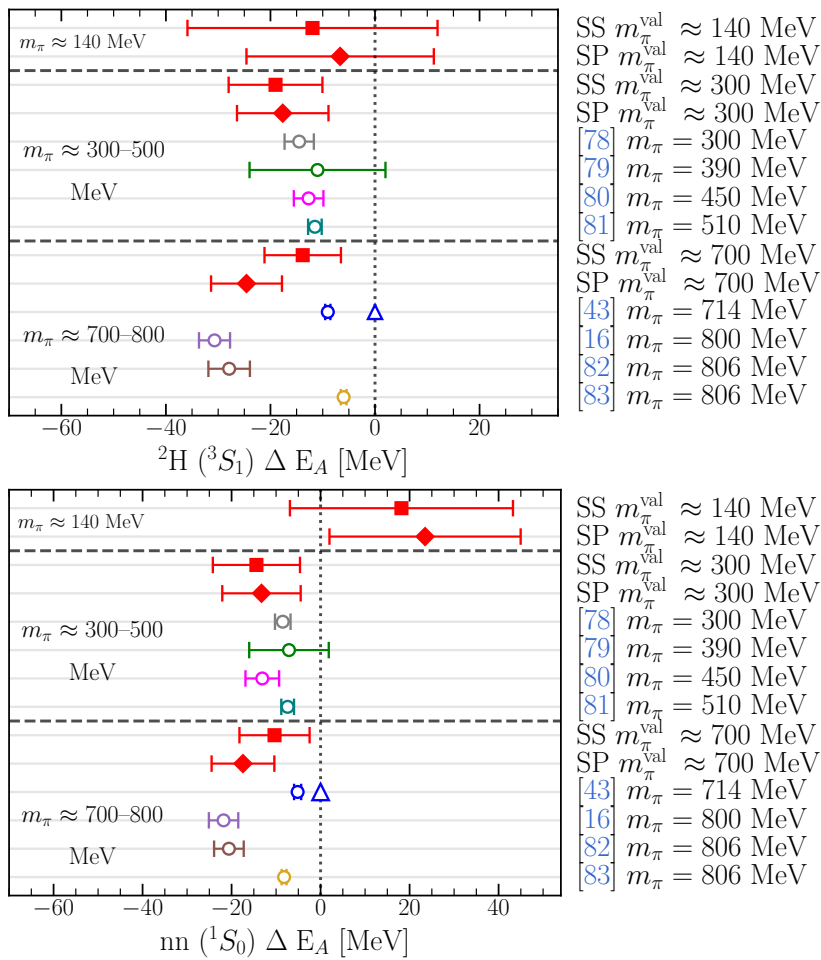
 
    \centering
    \begin{minipage}[b]{0.62\linewidth}
        \centering
        \resizebox{\linewidth}{!}{\input{Figs/deuteron_summary_plot.pgf}}
    \end{minipage}
    \begin{minipage}[b]{0.62\linewidth} 
        \centering
        \resizebox{\linewidth}{!}{\input{Figs/dineutron_summary_plot.pgf}} 
    \end{minipage}
    \caption{Summary of $\Delta E_A$ results for two-nucleon systems in the deuteron (top) and dineutron (bottom) channels from various lattice QCD calculations. Results are ordered from top to bottom by increasing valence pion mass. Results from this work are shown by filled red legends: diamonds denote the smeared–point (SP) and squares the smeared–smeared (SS) setups. This work uses physical-mass sea quarks and varying valence pion masses, while all previous studies used both heavy valence and sea pion masses.}
    \label{summaryNN}
\end{figure}
%

%

%\section{Results}

\section{\label{sec:ResultsforDeltaE}Results for \texorpdfstring{$\Delta E_A$}{Delta E\_A}}
Fig.~\ref{summaryNN} summarizes our extracted $\Delta E_A$ for deuteron and dineutron channels (filled markers), and compare them with previous lattice determinations. At heavier pion masses, $m_{\pi} \approx 700-800$~MeV, earlier lattice calculations reported relatively deeply bound states in both channels~\cite{NPLQCD2013,NPLQCD2017,CalLat2017}, while more recent analyses using enlarged operator bases and Lüscher’s formalism~\cite{LUSCHER1991531} find the systems to be unbound~\cite{NPLQCD2021,BaSc2025}. Our results place the lowest energy levels within $\sim10$~MeV of the two-nucleon threshold in both channels. A key distinction from earlier studies is that our calculations employ physical-mass sea quarks, whereas previous simulations used heavy sea quarks matched to the valence mass. At intermediate pion masses $m_{\pi} \approx 300$–$500$~MeV, existing lattice results are broadly consistent with each other. Our calculations likewise find the lowest energy levels in both channels below the threshold, with only mild variation compared with the heavier-mass regime. At the physical valance and sea pion masses, the deuteron channel remains consistent with a shallow bound state within uncertainties. In contrast, the dineutron channel shows a change in sign of the energy shift as the pion mass approaches its physical value, with the central value moving toward an unbound system. Although the uncertainties remain large, this behavior indicate that the qualitative difference between the two channels emerges near the physical quark masses. A full finite-volume amplitude analysis and enlarged operator basis will be required for a definitive determination.

\begin{figure}[htb!]
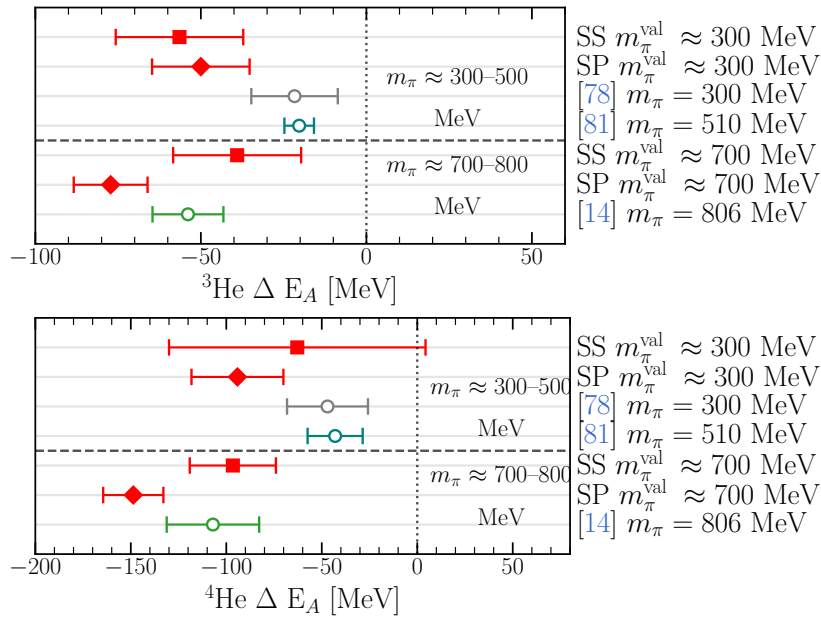
 
    \centering
    \begin{minipage}[b]{0.62\linewidth}
        \centering
        \resizebox{\linewidth}{!}{\input{Figs/helium3_summary_plot.pgf}}
    \end{minipage}
    \begin{minipage}[b]{0.62\linewidth} 
        \centering
        \resizebox{\linewidth}{!}{\input{Figs/helium4_summary_plot.pgf}} 
    \end{minipage}
    \caption{Summary of $\Delta E_A$ results for two-nucleon systems in the  $^3$He (top) and $^4$He (bottom) channels from various lattice QCD calculations. Results are ordered from top to bottom by increasing valence pion mass. Results from this work are shown by filled red legends: diamonds denote the smeared–point (SP) and squares the smeared–smeared (SS) setups. This work uses physical-mass sea quarks and varying valence pion masses, while all previous studies used both heavy valence and sea pion masses.}
    {\iffalse
    \caption{Summary of the results for energy shifts ($\Delta E_0$) for $^3$He ($\frac{1}{2}^+(\frac{1}{2})$, left panel) and $^4$He ($0^+(0)$, right panel) channels, as obtained in various lattice QCD calculations. Our results are shown by the filled symbols.}
    \fi}
    \label{summaryHe}
\end{figure}

In Fig.~\ref{summaryHe}, we summarize $\Delta E_A$ obtained in this work for $^3$He and $^4$He at $m_{\pi}^{\mathrm{val}} \approx 700$ and $300$~MeV together with existing lattice QCD results. Our results are consistent with the binding energies reported at similar pion masses in Ref.~\cite{NPLQCD2013}. At $m_{\pi} \approx 300$ MeV, the extracted energy shifts for $^3$He are somewhat smaller than those in Ref.~\cite{PACS2015}, while the $^4$He results agree within uncertainties. At the physical pion mass, the correlation functions for $^3$He and $^4$He are strongly noise dominated with our present statistics, preventing a reliable extraction of energy levels.

Both Figs.~\ref{summaryNN} and \ref{summaryHe} compare symmetric smeared–smeared (SS) and asymmetric smeared–point (SP) operator constructions. At $m_{\pi} \approx 700$ MeV the SS setup yields smaller (less negative) energy shifts than SP for all nuclei, indicating sensitivity to operator structure when several low-lying states are nearby (see Appendix. \ref{sec:SSvsSP}). At $m_{\pi} \approx 300$ MeV and near the physical point the SS and SP results are consistent within uncertainties. This work also represents the first application of the SS setup to three- and four-nucleon systems.

\begin{figure}[htb!]
    \includegraphics[width=0.6\linewidth]{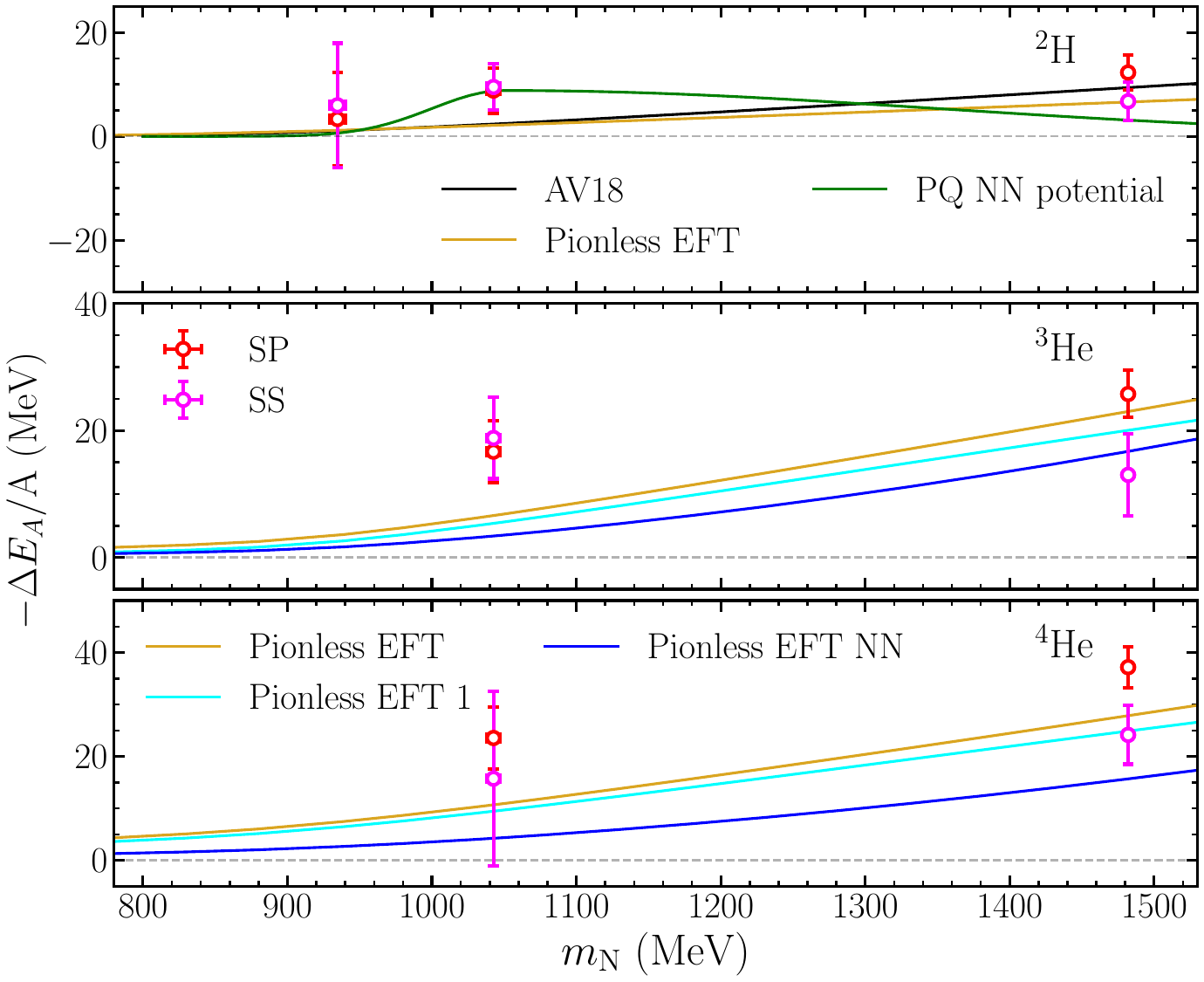}
    \caption{Lattice QCD results of the energy splitting $\Delta E_{A}$ as a function of nucleon mass $m_N$ compared with several low-energy descriptions for the deuteron (top), $^3$He (middle), and $^4$He (bottom). The curves correspond to the Argonne V18 phenomenological potential (AV18), next-to-leading order partially quenched nucleon–nucleon potential (PQ NN potential), leading-order pionless EFT (pionless EFT and pionless EFT1), and pionless EFT without three-nucleon interactions (pionless EFT NN).}
    \label{fig:compNP}
\end{figure}
\section{Comparison with low-energy descriptions}
The pion-mass dependence of nuclear binding is an important ingredient in several low-energy descriptions of nuclei. As a phenomenological baseline one may consider realistic nucleon–nucleon potentials such as the Argonne V18 (AV18) interaction~\cite{AV18}, whose parameters are fixed by experimental NN data. In pionless EFT~\cite{Kaplan_1996,Kaplan_1998,Kaplan2_1998,Bedaque1999,BEDAQUE2003589,Bedaque2002,Bedaque_2003}, pions are integrated out and nuclear interactions are represented by contact operators whose couplings encode short-distance QCD dynamics; quark-mass dependence then enters indirectly through hadron masses and the low-energy constants. In EFTs with explicit pions, the pion mass also affects the long-range interaction. In the Kaplan–Savage–Wise formulation \cite{Kaplan_1998} pion exchange is treated perturbatively around a nonperturbative contact interaction. The Beane-Bedaque-Savage-van Kolck (BBSvK) framework \cite{BEANE2002377} reorganizes this expansion about the chiral limit, retaining perturbative pions in the $^{1}S_{0}$ channel while resumming the chiral-limit tensor force at leading order in the coupled $^{3}S_{1}$–$^{3}D_{1}$ channel. In contrast, Weinberg-type chiral EFT \cite{RevModPhys.81.1773,machleidt2024recentadvanceschiraleft} constructs a pion-exchange potential that is iterated nonperturbatively and whose parameters encode the quark-mass dependence of nuclear forces. For mixed valence and sea quark masses, partially-quenched extension~\cite{PQPotential,BEANE200391} of the BBSvK explicit-pion EFT modify the long-distance meson-exchange interaction by introducing additional pseudo-Goldstone exchanges associated with the partially quenched chiral symmetry. These modifications affect both the $^{1}S_{0}$ and $^{3}S_{1}$–$^{3}D_{1}$ channels by altering the structure of the long-range potential, while preserving the BBSvK power counting.
 
Figure~\ref{fig:compNP} compares the lattice $\Delta E_A$ with representative low-energy descriptions. For the phenomenological approach we use the AV18 potential~\cite{AV18}, solving the coupled-channel Schr\"odinger equation in the $^{3}S_{1}$–$^{3}D_{1}$ channel while varying only the nucleon mass. For the  partially quenched nucleon–nucleon potential at next-to-leading order we solve the $^{3}S_{1}$–$^{3}D_{1}$ coupled-channel Schr\"odinger equation with an energy-dependent square-well regulator to control the short-distance divergence of the long-range potential. The long-range interaction arises from one- and two-pion exchange, while short-distance physics is encoded in local four-nucleon operators. The corresponding couplings are fixed by reproducing the physical deuteron binding energy and the lattice-determined binding energy at $m_{\pi}^{\mathrm{val}} \approx 300$~MeV. We also consider a pionless EFT at leading order~\cite{NuLattice} by fixing the two-nucleon contact interaction through the physical deuteron binding energy and the three-nucleon interaction through the physical binding energies of $^3$He (pionless EFT1) and $^4$He (pionless EFT), and varying the nucleon mass. Results without three-nucleon interactions (pionless EFT NN) are shown to illustrate how the lattice $\Delta E_A$ constrains the three-nucleon force. The lattice $\Delta E_A$ directly constrains how nuclear binding responds to changes in the quark masses and therefore informs the pion-mass dependence of low-energy nuclear interactions.

\begin{figure}[htb!]
  \centering
\includegraphics[width=0.62\linewidth]{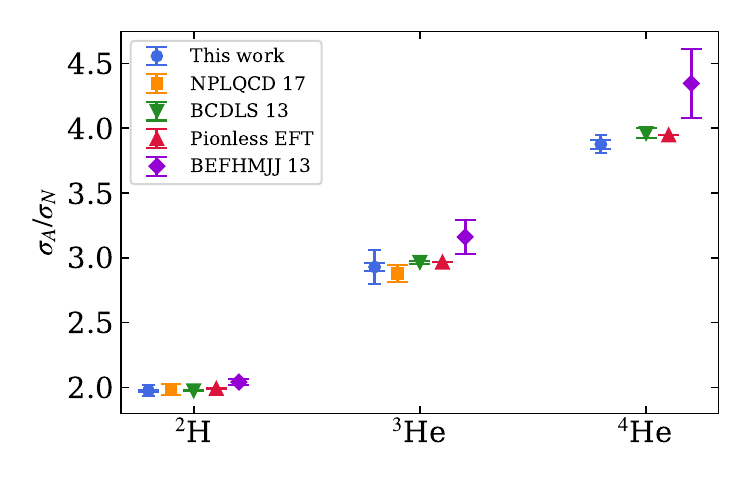}
  \caption{Comparison of nuclear to nucleon sigma-term ratios for ${}^2\mathrm{H}$, ${}^3\mathrm{He}$, and ${}^4\mathrm{He}$ from this work ((lattice QCD at physical quark masses) with previous lattice-QCD and EFT results: NPLQCD 17~\cite{Chang:2017eiq}, BCDLS 13~\cite{Beane:2013kca}, pionless EFT constrained in this work, and the EFT analysis of Ref.~\cite{Berengut:2013nh}. See text for details.}  
  \label{fig:sigmaA}
\end{figure}
\section{Nuclear sigma term}
The nuclear sigma term, $\sigma_A = {\partial m_A} / {\partial m_q}$, measures the contribution to the nuclear mass ($m_A$) arising from explicit chiral symmetry breaking due to the nonvanishing quark mass ($m_q$). From our results for $E_A(m_N)$, we constrain the ratio of nuclear to nucleon sigma terms ($\sigma_N = {\partial m_N} / {\partial m_q}$) using the Feynman–Hellmann theorem~\cite{Ohki:2008ff,Bali:2012qs,Bouchard:2016heu}
\begin{equation}
    \frac{\sigma_A}{\sigma_N} = 
    \frac{\partial E_A/\partial m_q}{\partial m_N/\partial m_q} =
    \frac{\partial E_A}{\partial m_N} \,.
\label{eq:rA_def}
\end{equation}

We determine ${\sigma_A}/{\sigma_N}$ from linear fits of $E_A(m_N)$ as a function of $m_N$. Fits are performed using only lattice data and also by including (${}^3\mathrm{He}$ and ${}^4\mathrm{He}$) or replacing (deuteron) the experimental physical-point nucleon and nuclear masses. We take the central values and statistical uncertainties from the fits including the physical-point input as our primary result, and assign the difference from the lattice-only fits as a systematic uncertainty. Statistical and systematic uncertainties are added in quadrature. The resulting ratios are shown in Fig.\ref{fig:sigmaA}, where the inner and outer error bars denote the statistical and total uncertainties, respectively. Within uncertainties, our results are consistent with earlier EFT-based analyses (BEFHMJJ 13)~\cite{Berengut:2013nh}, lattice QCD-based estimates (BCDLS 13)~\cite{Beane:2013kca}, lattice QCD results at $m_\pi \approx 800$~MeV (NPLQCD 17)~\cite{Chang:2017eiq}, and the corresponding pionless EFT results shown in Fig.~\ref{fig:compNP}. 
\begin{figure}[htb!]
  \centering
  \includegraphics[width=0.62\linewidth]{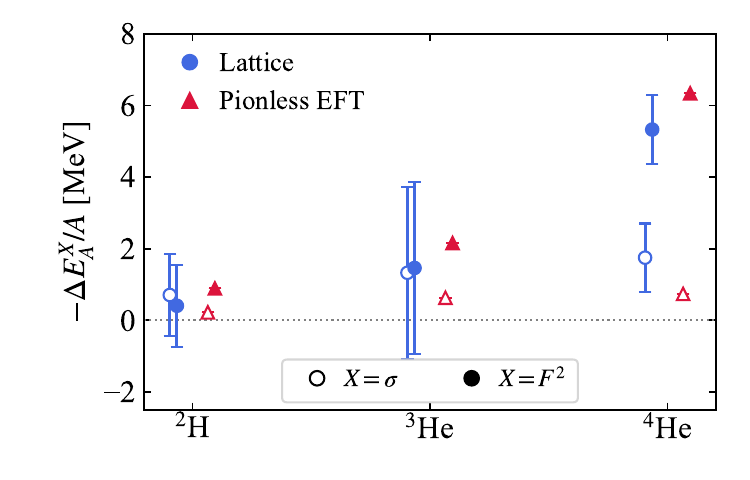}
  \caption{Decomposition of the binding energy per nucleon $\Delta E/A$ at $\overline{\mathrm{MS}}$=scale $\mu=2~\mathrm{GeV}$ into the quark-mass contribution, $\Delta E_{\sigma}$ (open symbols), and the gluonic trace-anomaly contribution, $\Delta E_{F^2}$ (filled symbols), for ${}^2\mathrm{H}$, ${}^3\mathrm{He}$, and ${}^4\mathrm{He}$. Circle symbols denote the lattice-QCD results, while triangle symbols correspond to the pionless-EFT benchmarks.}
  \label{fig:trace_decomp}
\end{figure}
\section{QCD origin of nuclear binding from trace anomaly}
To expose the QCD origin of nuclear binding, we start from the trace of the $\overline{\mathrm{MS}}$-renormalized QCD energy–momentum tensor at scale $\mu$,
\begin{equation}
    T^{\nu}_{\nu} =
    \frac{\beta(\mu)}{2g(\mu)}\,(F^2)_R +
    \left[ 1 + \gamma_m(\mu) \right] (m\bar\psi\psi)_R \,,
    \label{eq:trace_anomaly_prl}
\end{equation}
where $(F^2)_R(\mu)$ and $(m\bar\psi\psi)_R$ denote the renormalized gluonic and quark-mass operators, $g(\mu)$ is the renormalized strong coupling, $\beta(\mu)$ the QCD beta function, and $\gamma_m$ the quark-mass anomalous dimension. The forward matrix element of the EMT trace, $\langle A|T^\nu_{\ \nu}|A\rangle$, with $A=1$ corresponding to the nucleon ($N$) and $A>1$ to a nuclear bound state, yields the exact trace decomposition of the hadron mass~\cite{Hatta:2018sqd,Tanaka:2018nae,Tanaka:2022wzy,Tanaka:2025hhv}, $E_A = E_A^{\sigma}(\mu) + E_A^{F^2}(\mu)$, with $E_A^{\sigma}(\mu) = [ 1 + \gamma_m(\mu) ] \sigma_A$ and $E_A^{F^2}(\mu) = [\beta(\mu)/2g(\mu)] [\langle A|F^2|A\rangle / (2M_A)]$. Applying this relation to the nucleus and nucleon and subtracting on a per-nucleon basis yields a trace-anomaly decomposition of nuclear binding, $\Delta E_A = \Delta E_A^{\sigma}(\mu) + \Delta E_A^{F^2}(\mu)$, where 
\begin{equation}
\begin{split}
    & \Delta E_A^{\sigma}(\mu) =  \left[ 1 + \gamma_m(\mu) \right] \sigma_N 
       \left[ \frac{\sigma_A}{\sigma_N}-A \right] \,,
    \quad \mathrm{and} \\
    & \Delta E_A^{F^2}(\mu) = \frac{\beta(\mu)}{2g(\mu)} 
        \!\left[\frac{\langle A|F^2|A\rangle}{2M_A} -A \frac{\langle N|F^2|N\rangle}{2M_N}\right] \,.
\end{split}
\label{eq:deltaF2_sigma}
\end{equation}
This decomposition separates nuclear binding into a contribution from explicit chiral symmetry breaking, $\Delta E_A^{\sigma}(\mu)$, and a gluonic contribution associated with the trace anomaly, $\Delta E_A^{F^2}(\mu)$.

Using our lattice-determined $\sigma_A/\sigma_N$, with physical values $M_A$, lattice-QCD based $\sigma_N$ (43.7 MeV)~\cite{Agadjanov:2023efe}, and 3-loop $\overline{\mathrm{MS}}$ results for $\gamma_m(\mu)$ and $\beta(\mu)$~\cite{Tanaka:2018nae}, we determine $\Delta E_A^{\sigma}(\mu)$ and $\Delta E_A^{F^2}(\mu)$ around the deuteron mass scale of $\mu=2$~GeV. The results for the decomposition is shown in Fig.~\ref{fig:trace_decomp}. Repeating the analysis with the Roy--Steiner value of $\sigma_N$ (59.0 MeV)~\cite{Hoferichter:2023ptl} leads to the same qualitative pattern; those results are provided in the Appendix.\ref{sec:nuclear_trace_anomaly}. From ${}^2\mathrm{H}$ to ${}^4\mathrm{He}$, the increase in binding per nucleon is driven primarily by the growth of the gluonic contribution, with only mild variation in the sigma-term component. The distinct $A$-dependence admits a natural interpretation in analogy with the liquid-drop picture of nuclear binding. The quark-mass contribution per nucleon remains approximately constant, consistent with a bulk, mean-field–like component, while the growth of the gluonic contribution with $A$ indicates sensitivity to collective nuclear dynamics. Although the present nuclei are too light for a quantitative separation of volume and surface terms, this behavior suggests that the increase of nuclear binding with $A$ is driven primarily by gluonic many-body effects associated with the QCD trace anomaly. These results demonstrate the origin of nuclear binding from the QCD operator structure and show that light-nuclear binding is governed predominantly by gluonic dynamics.
\vspace*{-0.05in}

\section{Summary}
We have presented lattice QCD calculations of the finite-volume lowest energy levels of ${}^2\mathrm{H}$, $nn$, ${}^3\mathrm{He}$, and ${}^4\mathrm{He}$ nuclei, with physical sea quarks and valence quark masses spanning from heavy to near-physical pion masses. At the physical point, the two-nucleon channels exhibit the qualitative pattern of a shallow deuteron and an unbound dineutron within current uncertainties, while at heavier quark masses the systems become more deeply bound. A definitive determination of binding in these channels will require a full finite-volume amplitude analysis with an enlarged operator basis. Our results on the quark-mass dependence of the energies provide first-principles constraints on low-energy nuclear interactions and is broadly consistent with expectations from phenomenological potentials and effective field theory descriptions.

From the quark-mass response of nuclear energies, we determine nuclear sigma terms and quantify the role of explicit chiral symmetry breaking in nuclear binding. The results support approximate additivity of the sigma terms with nucleon number, with only small non-additive contributions within current precision.

Using these inputs, we perform a trace-anomaly–based decomposition of nuclear binding. We find that the quark-mass contribution to the binding energy per nucleon is subleading, while the dominant contribution arises from the gluonic component associated with the QCD trace anomaly. The distinct mass-number dependence of these contributions suggests that the growth of nuclear binding with mass number is driven primarily by gluonic many-body dynamics, with quark-mass effects providing a smaller correction. These results establish a direct connection between nuclear binding and its QCD origin in the trace anomaly, and provides a direction to investigate it with quantitative precision in future.

\vspace*{-0.05in}

\section{Acknowledgement}
This material is based upon work supported by the U.S.~Department of Energy, Office of Science, Office of Nuclear Physics through Contract No.~DE-SC0012704, and within the frameworks of Scientific Discovery through Advanced Computing (SciDAC) award Fundamental Nuclear Physics at the Exascale and Beyond. This work is supported by the Department of Atomic Energy, Government of India, under Project Identification Number RTI 4012.

This research used awards of computer time provided by the Oak Ridge Leadership Computing Facility, which is a DOE Office of Science User Facility supported under Contract DE-AC05-00OR22725, and the National Energy Research Scientific Computing Center, a DOE Office of Science User Facility
supported by the Office of Science of the U.S. Department of Energy under Contract No. DE-AC02-05CH11231 using NERSC award
NP-ERCAP0032114 and NP-ERCAP0036132. This work also used the computing facilities of the 
Department of Theoretical Physics, TIFR, Mumbai. 

This work used PyQUDA~\cite{Jiang:2024lto} and SIMULATeQCD~\cite{HotQCD:2023ghu} software packages. NM and DC thank Piyush Srivastava for valuable discussions on the development of contraction codes.

\clearpage
\appendix

\section{Interpolating operators for single nucleon and light nuclei}\label{sec:op_const}
In this section, we outline the operators used in this study to compute the two-point correlation functions for both the single nucleon and the light nuclei investigated. For the light nuclei, we employ the operators introduced in Ref.~\cite{PACS2012}, which we discuss below. For single nucleons, the following interpolating operators are employed:
\begin{eqnarray}
    \mathcal{P}_{\alpha}(\mathbf{x},t) &\equiv& (P_+)_{\alpha\eta}\epsilon_{abc} u_a^{\beta}(\mathbf{x},t)(C\gamma_5)_{\beta\gamma}d_b^{\gamma}(\mathbf{x},t) u_c^{\eta}(\mathbf{x},t), \label{eq:proton_op} \\
    \mathcal{N}_{\alpha}(\mathbf{x},t) &\equiv& (P_+)_{\alpha\eta} \epsilon_{abc}d_a^{\beta}(\mathbf{x},t)(C\gamma_5)_{\beta\gamma}u_b^{\gamma}(\mathbf{x},t)d_c^{\eta}(\mathbf{x},t). \label{eq:neutron_op}
\end{eqnarray}
Here, $\mathcal{P}_\alpha(\mathbf{x},t)$ and $\mathcal{N}_\alpha(\mathbf{x},t)$ denote the proton and neutron interpolating fields, respectively. In Eqs.~(\ref{eq:proton_op}) and (\ref{eq:neutron_op}), $ C = i\gamma_2\gamma_4 $ is the charge conjugation matrix, and $ P_+ = \frac{1}{2}(\mathrm{I} + \gamma_4) $ is the positive-parity projection operator. Using these nucleon operators as the building block, we construct the interpolating fields for the two-nucleon channels $^3S_1 (NP)$ and $^1S_0 (NN)$, as well as for the three- and four-nucleon states, ${}^3$He and ${}^4$He, respectively as below:
\begin{eqnarray}
    \mathcal{O}_{^3S_1}^i(\mathbf{x},t) &=& \frac{1}{\sqrt{2}} \left( \mathcal{P}_{\alpha}(\mathbf{x},t) (C\gamma_i)_{\alpha\beta} \mathcal{N}_\beta(\mathbf{x},t)\right.\nonumber\\
    &&\left.- \mathcal{N}_{\alpha}(\mathbf{x},t) (C\gamma_i)_{\alpha\beta} \mathcal{P}_\beta(\mathbf{x},t) \right), \label{NP_op} \\
    \mathcal{O}_{^1S_0}(\mathbf{x},t) &=& \mathcal{N}_{\alpha}(\mathbf{x},t) (C\gamma_5)_{\alpha\beta} \mathcal{N}_\beta(\mathbf{x},t), \label{NN_op} \\
    \mathcal{O}^\uparrow_{^3{\mathrm{He}}}(\mathbf{x},t) &=& \frac{1}{\sqrt{6}}\left(\mathcal{P}_\downarrow\mathcal{N}_\uparrow\mathcal{P}_{\uparrow} - \mathcal{P}_\uparrow\mathcal{N}_\uparrow\mathcal{P}_{\downarrow} + \mathcal{N}_\uparrow\mathcal{P}_\uparrow\mathcal{P}_{\downarrow}\right.\nonumber\\&& -\mathcal{N}_\uparrow\mathcal{P}_\downarrow\mathcal{P}_{\uparrow}+\mathcal{P}_\uparrow\mathcal{P}_\downarrow\mathcal{N}_{\uparrow}-\mathcal{P}_\downarrow\mathcal{P}_\uparrow\mathcal{N}_{\uparrow}\left.\right), \label{He3_op} \\
    \mathcal{O}_{^4{\mathrm{He}}} &=& \frac{1}{\sqrt{2}} \left( \bar{\mathcal{I}}\otimes \mathcal{S} - \mathcal{I}\otimes \bar{\mathcal{S}} \right). \label{He4_op}
\end{eqnarray}

In Eq.~(\ref{He4_op}), $\mathcal{S}$ and $\bar{\mathcal{S}}$ denote the spin wave functions, and are defined as (Refs.~\cite{PhysRev.158.907,PACS2015,Chakraborty:2024oym}):
\begin{eqnarray}
    \mathcal{S} &=& \frac{1}{2} \left( [\uparrow \downarrow \uparrow \downarrow] + [\downarrow \uparrow \downarrow \uparrow] - [\uparrow \downarrow \downarrow \uparrow] - [\downarrow \uparrow \uparrow \downarrow] \right), \\
    \bar{\mathcal{S}} &=& \frac{1}{\sqrt{12}} \left( [\uparrow \downarrow \uparrow \downarrow] + [\downarrow \uparrow \downarrow \uparrow] + [\uparrow \downarrow \downarrow \uparrow] + [\downarrow \uparrow \uparrow \downarrow] \right. \nonumber \\
    &&\left. - 2 [\uparrow \uparrow \downarrow \downarrow] - 2 [\downarrow \downarrow \uparrow \uparrow] \right).
\end{eqnarray}
Here, $\uparrow$ and $\downarrow$ represent the spin-up and spin-down components of the nucleons. The isospin wave functions $\mathcal{I}$ and $\bar{\mathcal{I}}$ are constructed analogously by replacing the spin doublet $(\uparrow, \downarrow)$ with the isospin doublet $(\mathcal{P}, \mathcal{N})$. The interpolating operators defined in Eqs.~(\ref{NP_op})–(\ref{He4_op}) are constructed to have well-defined spin and isospin quantum numbers corresponding to the physical multi-nucleon systems they represent. 

\section{Numerical Setup\label{numerical_setup}}
We employ  $N_f=2+1$ flavor lattice QCD ensemble generated by HotQCD collaboration \cite{Follana:2006rc, HotQCD:2014kol, Bazavov:2019www} using Highly Improved Staggered Quark (HISQ) dynamics \cite{HISQaction}. The lattice dimension is $64^3\times 64$ with a lattice spacing of $a=0.076$ fm. The sea light quark masses are adjusted to produce a pion mass of $m_{\pi}^{\text{sea}}=140$ MeV while the sea strange quark mass is tuned to reproduce $m_{\eta_{s\overline{s}}}=686$ MeV. For the valence sector, we employ a tree-level tadpole improved Wilson--Clover action with one level of hypercubic (HYP) smearing \cite{HYPSmearing} applied to the background gauge fields. The clover coefficient $c_{\text{sw}}$ is set to the tree-level tadpole-improved value computed using the fourth root of the average plaquette $u_0$, yielding $c_{\text{sw}}=u_0^{-3/4}$, which is $1.0372$ in our case. The calculation is performed for three sets of values for the kappa $\kappa=0.12642,\,0.12639,\,0.12547$, which correspond to valence pion masses of $140$, $295$ and $695$ MeV, respectively.

On this lattice ensemble, we compute the two-point correlation function, $C_{2\text{pt}}(\tau)$, of interpolating operators that carry the isospin, spin and parity quantum numbers of ground state of the desired light nuclei, with the operators inserted at a Euclidean time separation $\tau$,
\begin{equation}
C_{2\text{pt}}(\tau = t_f - t_i) =
\sum_{\mathbf{x}_f} \langle \Omega \vert\, \mathcal{O}(\mathbf{x}_f, t_f)\, \overline{\mathcal{O}}(\mathbf{x}_i, t_i)\, \vert \Omega \rangle .
\label{eq:C2pt}
\end{equation}
The finite-volume energy levels of light nuclei states can be extracted from the exponential falloff of $C_{2\text{pt}}(\tau)$ with Euclidean time $\tau$. For deuteron, dineutron, helium-3 and helium-4, we employ the operators, outlined in Sec.(\ref{sec:op_const}). To compute this two-point correlation function, one first numerically calculates the quark propagators by inverting the Dirac operator on each gauge configuration, then performs all Wick contractions by tying together the propagators, and finally averages the resulting correlation functions over the gauge ensemble. To generate the quark propagators, we apply a multigrid algorithm~\cite{Multigrid} to invert the Wilson--Dirac operator using \texttt{PyQUDA}~\cite{PyQUDA}, a Python wrapper for the \texttt{QUDA} software suite~\cite{CLARK20101517,10.1145/2063384.2063478, clark2016acceleratinglatticeqcdmultigrid}. The quark sources are constructed in Coulomb gauge with a Gaussian profile of radius $\sim 0.684$ fm. We utilize asymmetric smeared-source/point-sink (SP) and symmetric smeared-source/smeared-sink (SS) setups to compute the two-point correlation functions. To increase statistics, multiple source locations are used on each gauge configuration. Table~\ref{tab:params} lists the number of configurations and the number of sources per configuration for each of the three $\kappa$ values, for both SP and SS smearing.To compute the two-point correlation functions of multinucleon systems, we use a scalable MPI-enabled implementation of the contraction algorithm described in~\cite{Chakraborty:2024oym}, which leverages GPU acceleration.

\begin{table}[h!]
    \centering
    \renewcommand\arraystretch{1.2}  % Increase the height of each row
    \addtolength{\tabcolsep}{3pt}    % Increase separation between columns
    \begin{tabular}{|c|c|c|c|c|}
     \hline\hline 
      $\kappa$ & $m_{\pi}^{\text{val}}$ (MeV) & $m_{\text{N}}$ (MeV)& smearing & $n_{\text{src}}\times n_{\text{cfg}}$ \\ \hline
      \multirow{2}{*}{$0.12547$} & \multirow{2}{*}{$695(1)$} & \multirow{2}{*}{$1478(2)$} & SP & $64\times660$ \\ \cline{4-5}
      & & & SS & $256\times 700$ \\
      \hline
      \multirow{2}{*}{$0.12639$} & \multirow{2}{*}{$295(1)$} & \multirow{2}{*}{$1045(3)$} & SP & $256\times 700$ \\ \cline{4-5}
      & & & SS & $512\times 700$ \\ \hline
      \multirow{2}{*}{$0.12642$} & \multirow{2}{*}{$140(1)$} & \multirow{2}{*}{$932(5)$} & SP & $700\times 1300$ \\ \cline{4-5}
      & & & SS & $700\times 1300$ \\ \hline
    \end{tabular}
   \caption{Details of the lattice parameters $\kappa$, valence pion and nucleon masses, 
and statistics used for the different smearing setups employed in this study. 
Here, `SP' denotes the asymmetric smeared--point setup, while `SS' denotes 
the smeared--smeared setup.}

    \label{tab:params}
\end{table}

\section{Extraction of finite volume energy levels\label{sec:DetailsEnergyExtraction}}
We extract the lowest-lying finite-volume energy levels of the nuclei of interest from the exponential decay of the corresponding two-point correlation functions ($C_{\text{2pt}}(\tau)$) at large Euclidean time separations ($\tau$) between the source and the sink. To reliably determine these energies, we employ two complementary analysis procedures for the correlation functions. Specifically, we use both standard multi-state fittings and the recently proposed transfer matrix generalized eigenvalue problem (TGEVP) method~\cite{Chakraborty:2024scw} to extract the energy levels. The results obtained from these two approaches are in excellent agreement within uncertainties, demonstrating the robustness of our energy extraction. Below, we describe each of these analysis strategies in detail.

\subsection{Multi-State Fit \label{subsec:MultiStateFit}}
The functional form used to fit the two-point correlation functions of individual light nuclei and the single nucleon is given by,
\begin{eqnarray}
C^A_{\text{2pt}}(\tau)=\sum_{n=0}^{n_{\text{max}}}\vert Z_A^n\vert^2 \exp(-\tau E_A^n),
\label{eq:fitform}
\end{eqnarray}
which follows from the spectral decomposition of two-point correlation functions. In Eq.~\ref{eq:fitform}, it is assumed that at sufficiently large source–sink time separations, the correlation function is well described by the contributions from the lowest-lying $n_{\text{max}}$ states, with energies $E_A^n$ and overlap factors $Z_A^n$. Our primary interest is in extracting the ground-state energy, $E_A \equiv E_A^0$. To this end, we perform correlated fits to the two-point correlation functions of light nuclei and the single nucleon individually, using the above fit ansatz within a bootstrap framework. From the extracted ground-state energies, we compute the finite-volume energy shifts of the light nuclei, defined as $\Delta E_A \equiv E_A - A m_N$, where $A m_N$ denotes the lowest non-interacting $A$-nucleon threshold. All stages of the analysis are carried out within the bootstrap procedure to account for statistical correlations. To assess the stability of the fits, we vary the fit ranges within the effective mass plateau and use the resulting spread to estimate the associated fit-window systematic uncertainties. 

\begin{figure}[h!]
    \centering
    \includegraphics[width=0.45\linewidth]{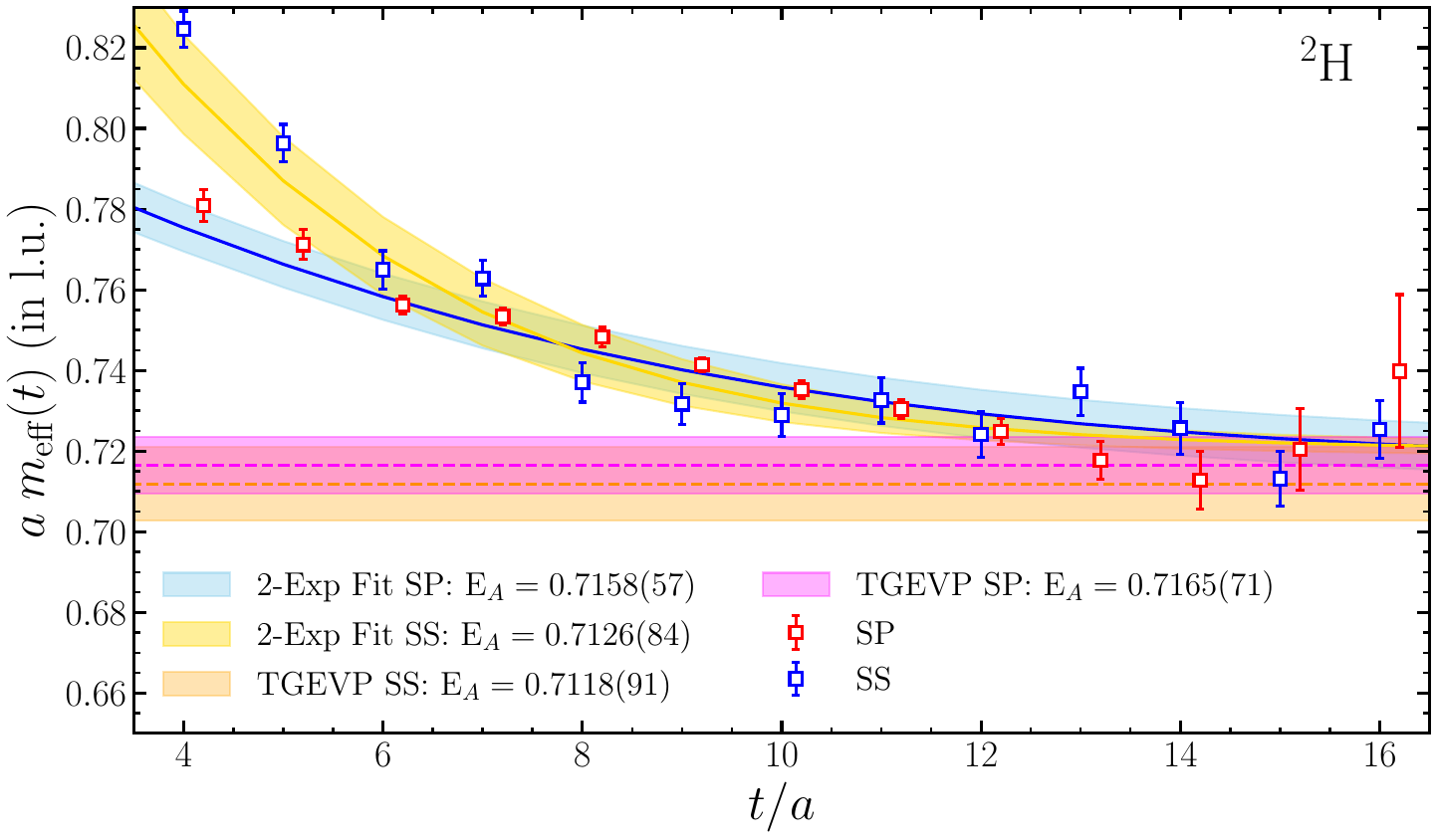}
    \includegraphics[width=0.45\linewidth]{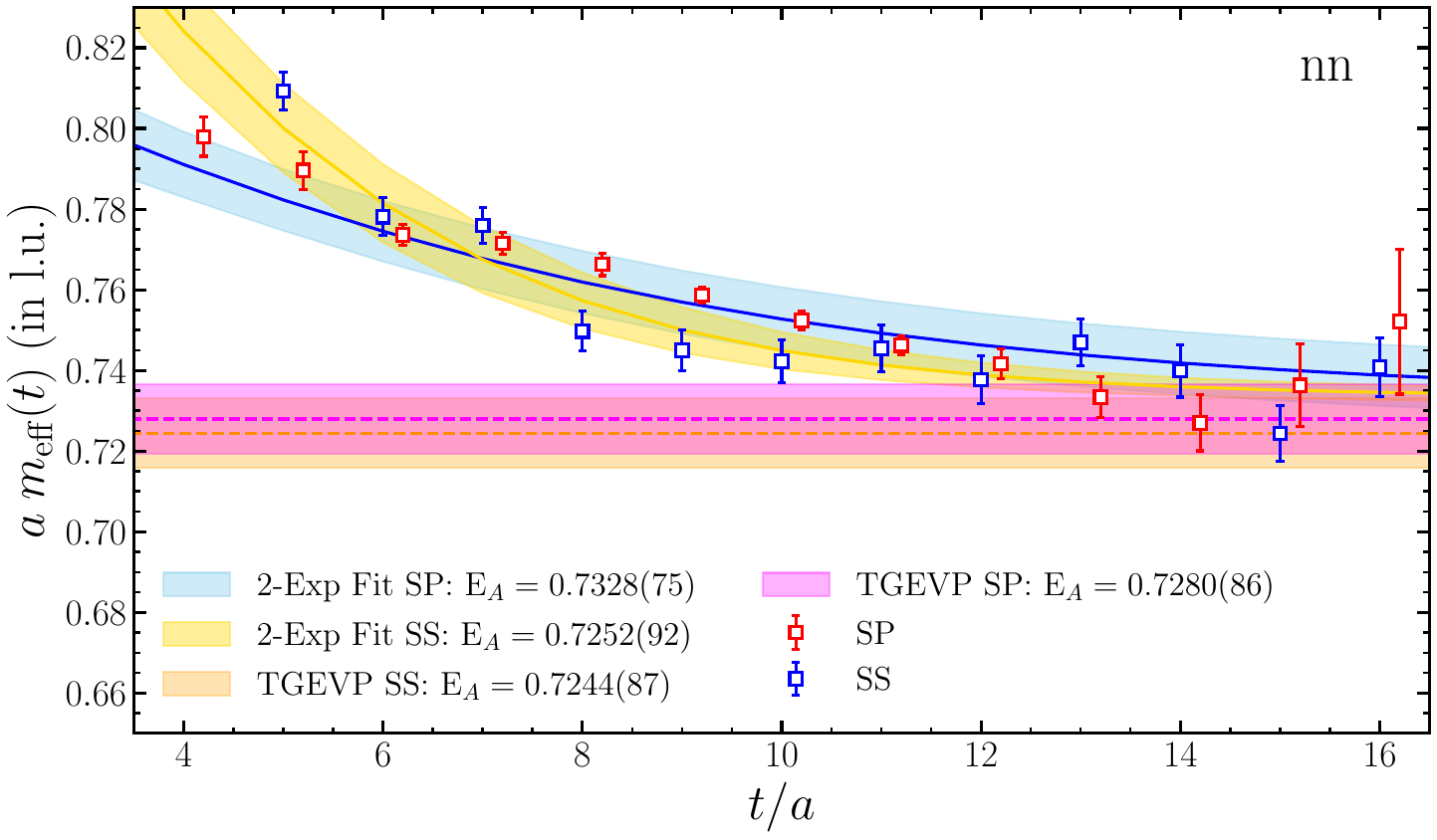}
    \includegraphics[width=0.45\linewidth]{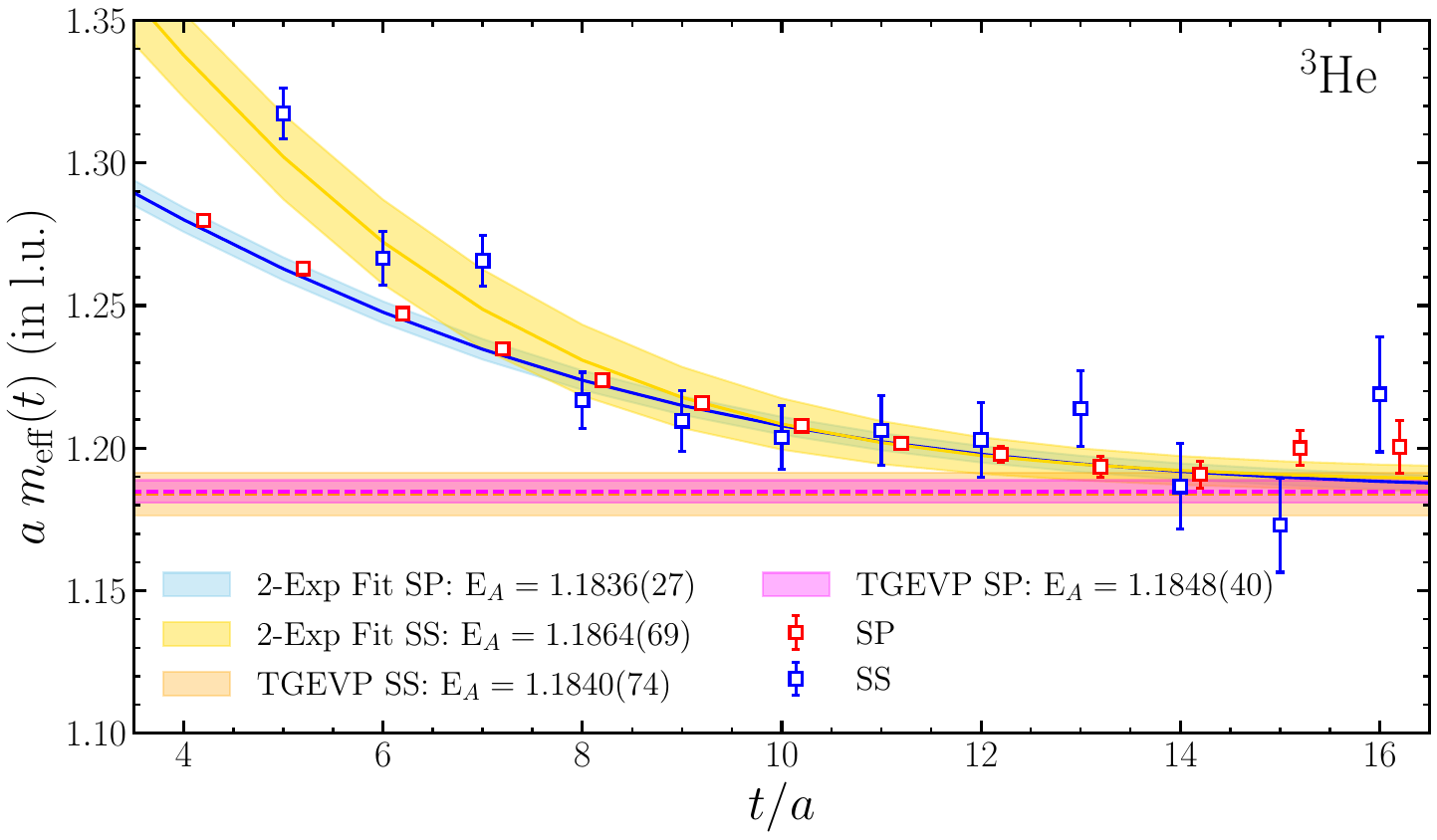}
    \includegraphics[width=0.45\linewidth]{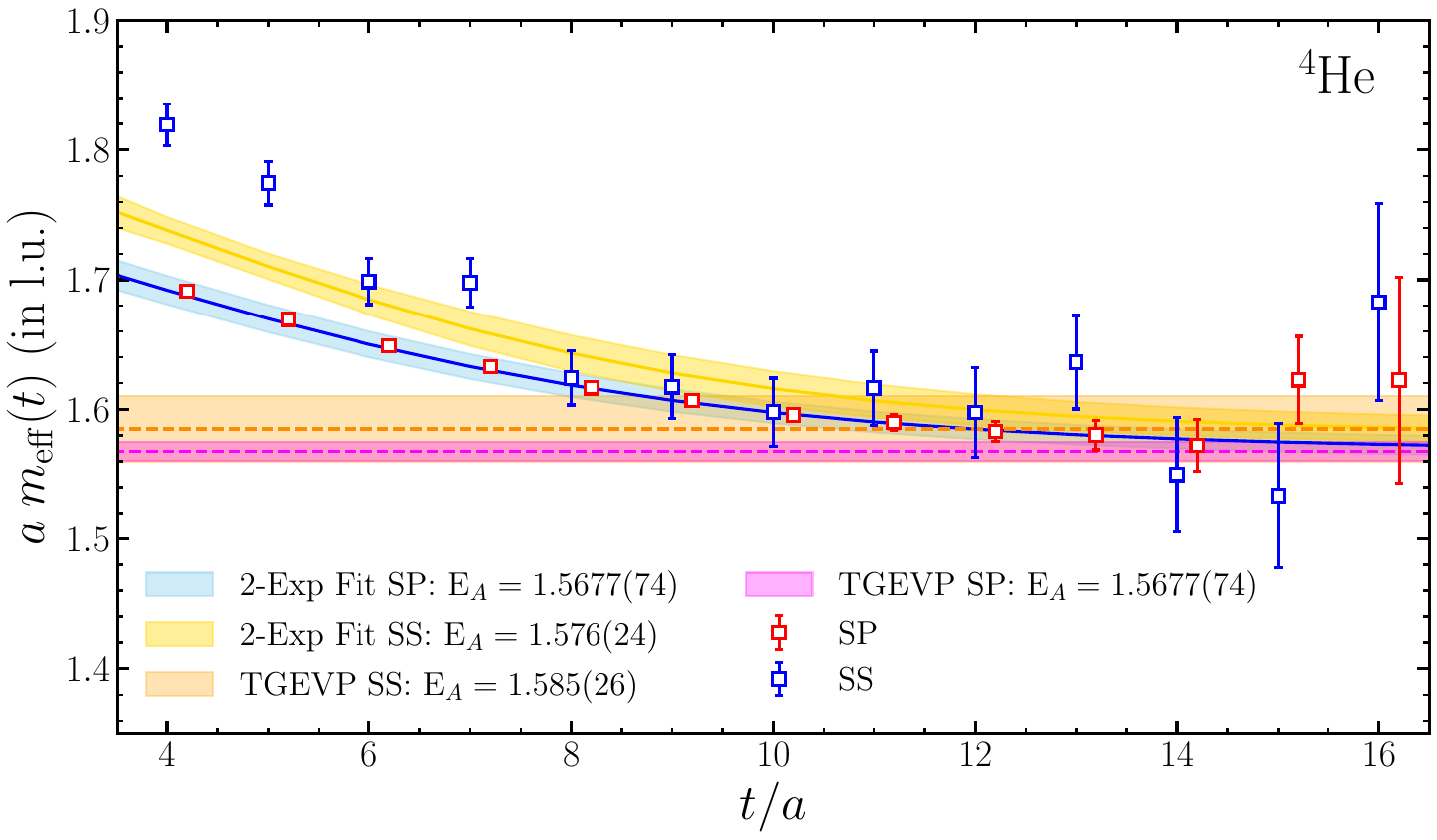}
    \caption{Effective masses of the deuteron ($^2$H), dineutron (nn), $^3$He, and $^4$He for both SS and SP setups. For $^2$H and nn, results at a valence pion mass of $m_{\pi} = 140~\mathrm{MeV}$ are shown, while for $^3$He and $^4$He the results correspond to $m_{\pi} \approx 300~\mathrm{MeV}$. The blue and yellow bands represent the two-state fit results for the SP and SS setups, respectively, with their associated uncertainties. The energy levels extracted using the TGEVP method are overlaid as bands, obtained from SS (orange) and SP (magenta), with $1\sigma$ uncertainties.}

    \label{fig:effmass}
\end{figure}

In Fig.~\ref{fig:effmass}, we present the effective masses for the deuteron ($^2$H), dineutron (nn), $^3$He, and $^4$He obtained from both SS and SP setups. For the two-nucleon systems ($^2$H and nn), the results correspond to a valence pion mass of $m_{\pi}^{\mathrm{val}} = 140~\mathrm{MeV}$, while for the three- and four-nucleon systems ($^3$He and $^4$He) we show results at $m_{\pi}^{\mathrm{val}} \approx 300~\mathrm{MeV}$.
The bands indicate the energy levels extracted from two-state fits, with the SP (SS) results shown in blue (yellow). For comparison, we overlay the energy levels obtained using the TGEVP method as bands with $1\sigma$ uncertainties, derived from SS (orange) and SP (magenta) correlators. We observe excellent agreement between the two-state fit and TGEVP determinations across all channels, indicating a robust and reliable extraction of the ground-state energies. In addition, the consistency between the SP and SS setups within uncertainties suggests that systematic effects associated with operator choices are well under control.

\subsection{Transfer Matrix GEVP (TGEVP) \label{subsec:TGEVP}}
We also employ the recently developed lattice transfer matrix formalism in a Krylov subspace ~\cite{Lanczos,BlockLanczos,Ostmeyer:2024qgu,Chakraborty:2024scw}, as a complementary cross-check for extracting energy levels from the two-point correlation functions. In this approach, one computes 
the eigenvalues $\lambda_n$ of the lattice transfer operator
$
\hat{\mathcal{T}} \equiv e^{-a\hat{\mathcal{H}}}
$
projected onto the Krylov subspace,
\[
\mathcal{K}_m \equiv \left\{\, 
\mathcal{T}\mathcal{O}|\Omega\rangle,\,
\mathcal{T}^2 \mathcal{O}|\Omega\rangle,\,
\dots,\,
\mathcal{T}^m \mathcal{O}|\Omega\rangle
\right\}, \, 
\lambda_n = e^{-E_n}.
\]
In practice, one generally solves the generalized eigenvalue problem (GEVP) constructed from time-shifted two-point correlation functions, and treats spurious eigenvalues arising from statistical fluctuations using various prescriptions. In this work, we follow the TGEVP procedure of Ref.~\cite{Chakraborty:2024scw}, which provides a robust framework for systematically handling such spurious modes. Specifically, we first perform a bootstrap-based estimation of the eigenvalue density using an adaptive kernel density estimator (KDE). From this distribution, we identify resolvable states by extracting the centers and widths of the spectral peaks. These peak characteristics are then used to define threshold criteria for eliminating spurious eigenvalues. After removing these unphysical modes, standard statistical techniques are applied to determine the mean values and uncertainties of the extracted energy levels. We cross-check the extracted energy levels from multi-state fits with TGEVP results to assess the robustness of the energy extraction.

\begin{figure}[h!]
    \centering    \includegraphics[width=\linewidth]{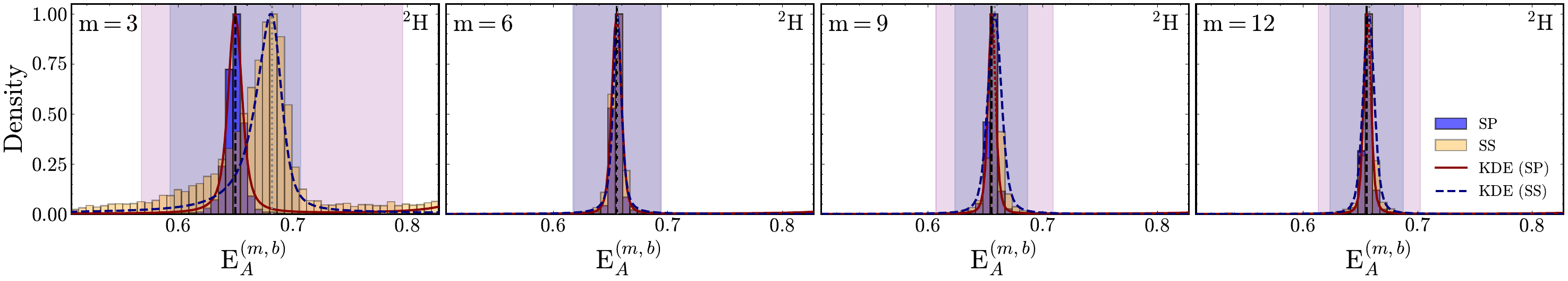}
    \centering    \includegraphics[width=\linewidth]{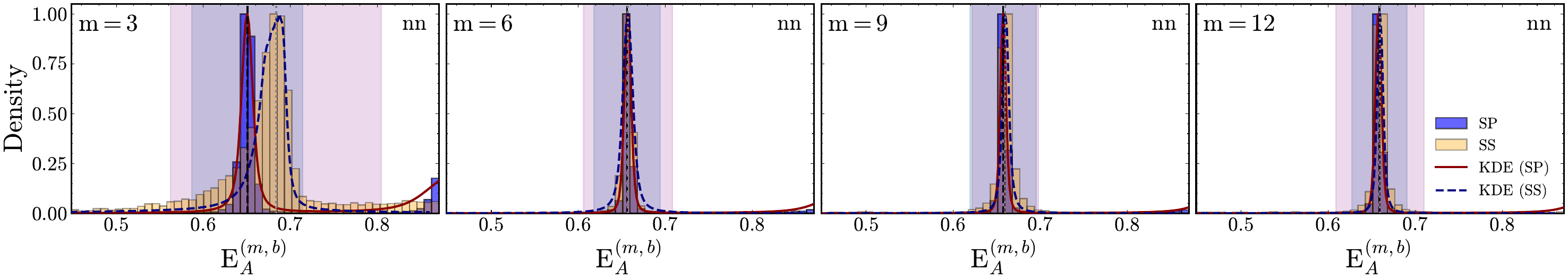}
    \caption{Bootstrapped eigenvalue distributions for light nuclei $^2$H (top) and nn (bottom) at valance $m_{\pi} \approx 140~\mathrm{MeV}$, for both SS and SP setups, obtained at various iterations of the TGEVP, are shown as histograms. The corresponding density estimates using adaptive kernel density estimation (KDE) are overlaid as a solid red curve for SP and a dashed black curve for SS. The sky-blue and purple bands indicate the threshold regions defined by $3\times\mathrm{FWHM}$ around the peak centers of the SP and SS distributions, respectively.}
    \label{fig:eig_distr_2N}
\end{figure}

\begin{figure}[h!]
    \centering    \includegraphics[width=\linewidth]{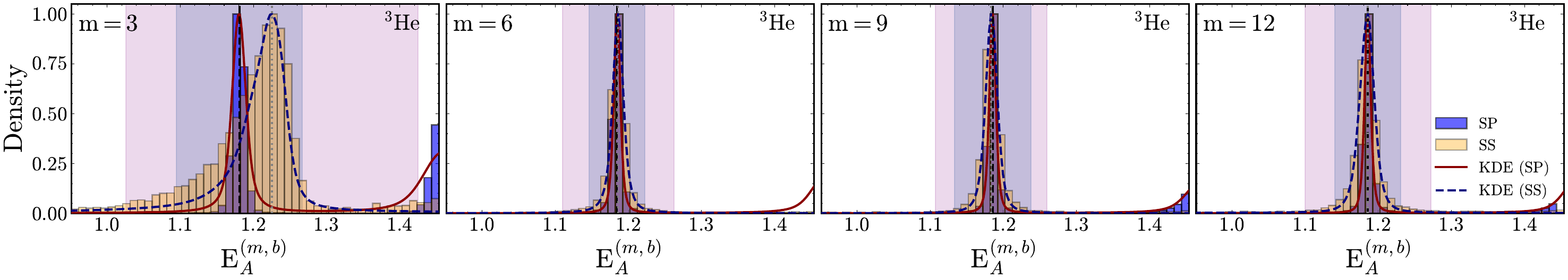}
    \centering    \includegraphics[width=\linewidth]{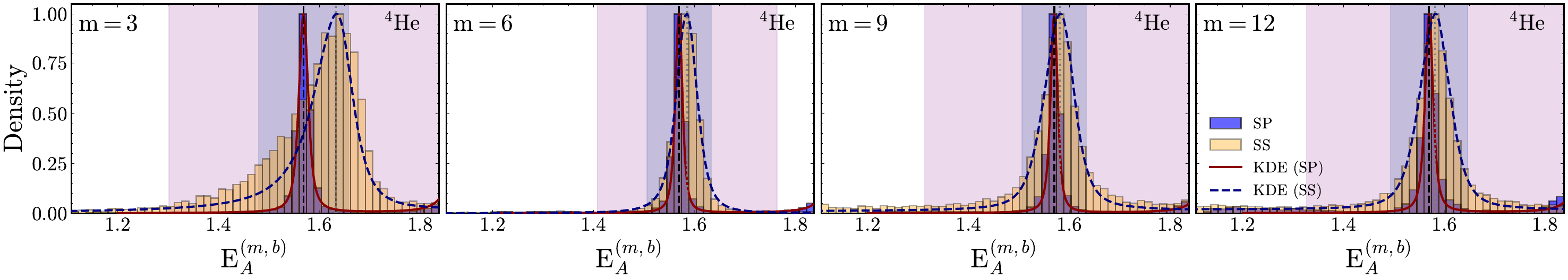}
    \caption{Bootstrapped eigenvalue distributions for light nuclei $^3$He (top) and $^4$He (bottom) at valance  $m_{\pi} \approx 300~\mathrm{MeV}$, for both SS and SP setups, obtained at various iterations of the TGEVP, are shown as histograms. The corresponding density estimates using adaptive kernel density estimation (KDE) are overlaid as a solid red curve for SP and a dashed black curve for SS. The sky-blue and purple bands indicate the threshold regions defined by $3\times\mathrm{FWHM}$ around the peak centers of the SP and SS distributions, respectively.}
    \label{fig:eig_distr_He}
\end{figure}

The procedure is illustrated in Figs.~\ref{fig:eig_distr_2N} and \ref{fig:eig_distr_He}, where we show the bootstrapped eigenvalue distributions for $^2$H and nn at $m_{\pi} \approx 140~\mathrm{MeV}$, and for $^3$He and $^4$He at $m_{\pi} \approx 300~\mathrm{MeV}$ (from top to bottom, respectively), for both SS and SP setups across several iterations of the TGEVP. The histograms represent the distribution of eigenvalues obtained from bootstrap samples, while the overlaid adaptive KDE provides a smooth estimate of the underlying spectral density, shown as a solid red curve for SP and a dashed black curve for SS. Distinct peaks in these distributions correspond to physical energy levels, whereas broad or irregular structures typically signal contributions from statistical noise or unresolved states. The sky-blue and purple bands indicate the threshold regions defined by $3\times \mathrm{FWHM}$ around the peak centers of the SP and SS distributions, respectively, and serve as selection windows for isolating physically meaningful eigenvalues. Eigenvalues falling outside these regions are systematically identified as spurious and removed from further analysis.

\begin{figure}[h!]
   \includegraphics[width=0.24\linewidth]{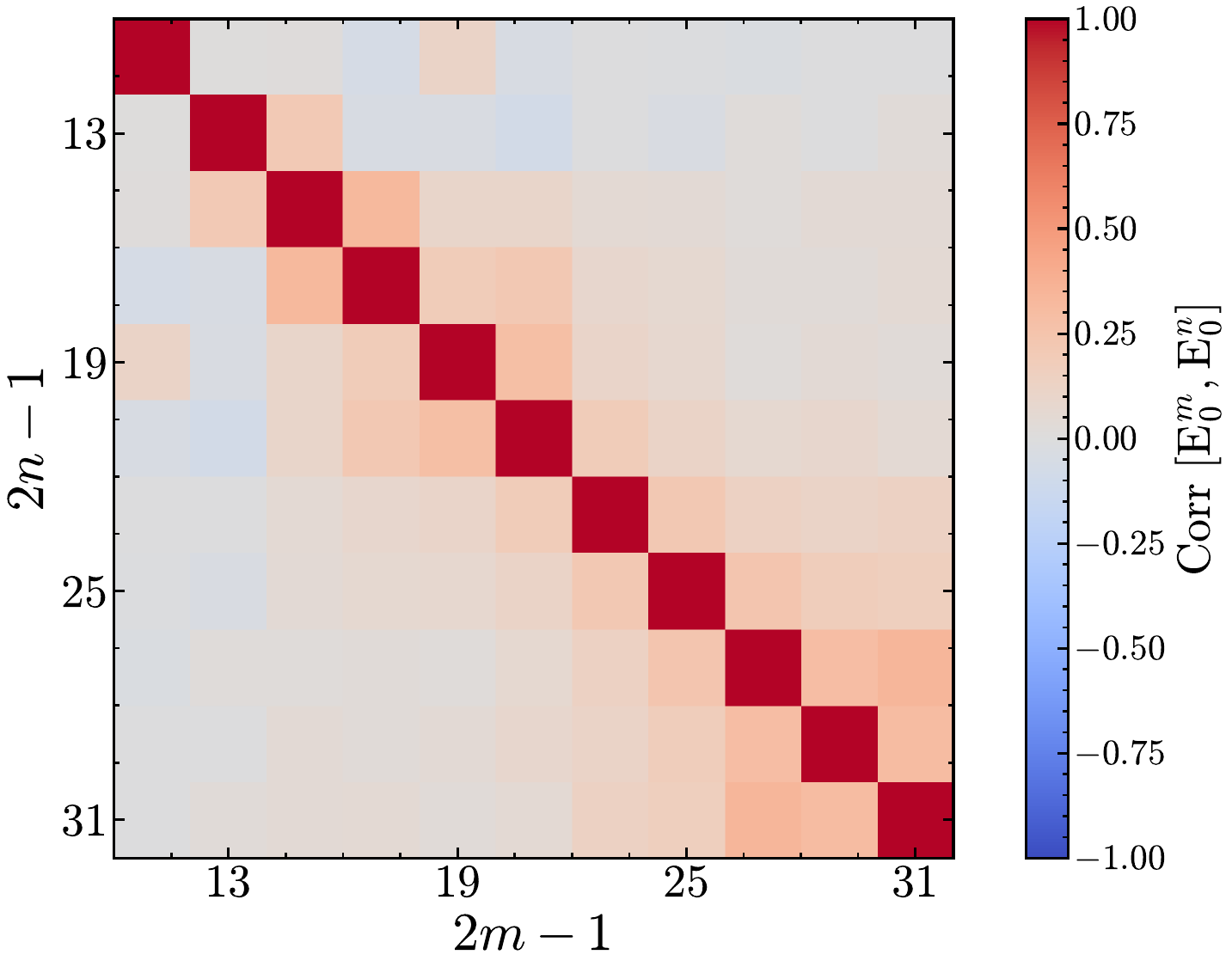}
   \includegraphics[width=0.24\linewidth]{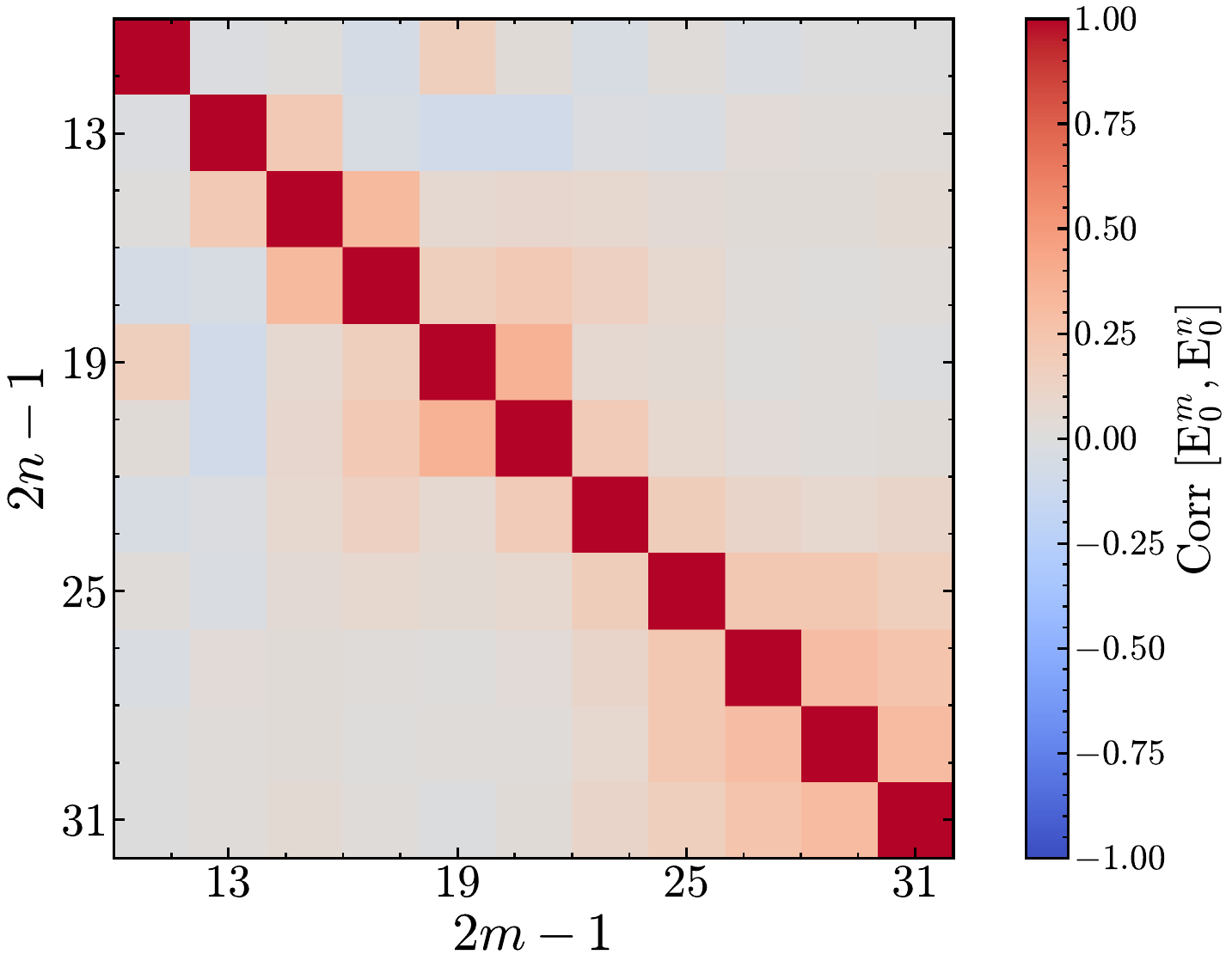}
   \includegraphics[width=0.24\linewidth]{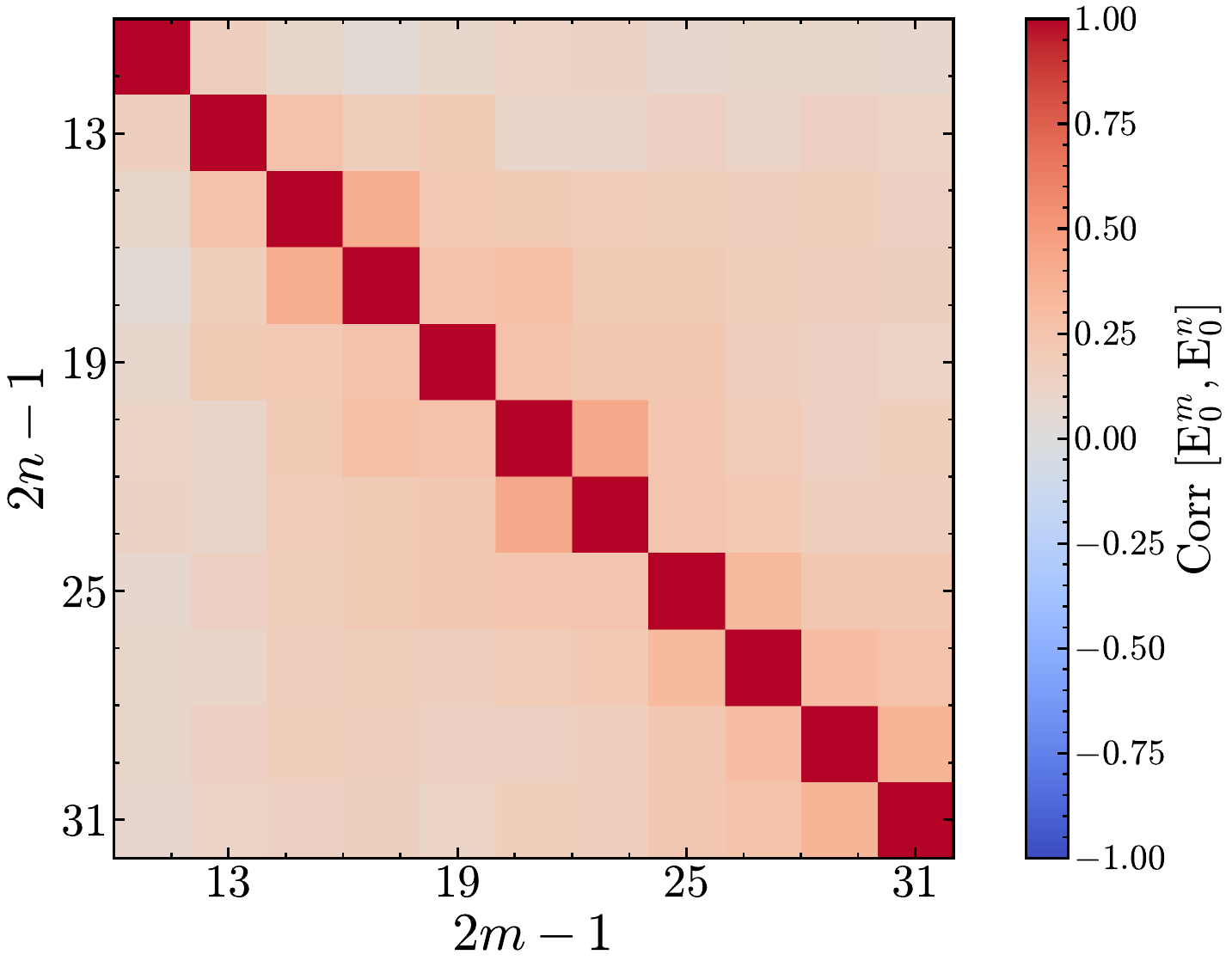}
   \includegraphics[width=0.24\linewidth]{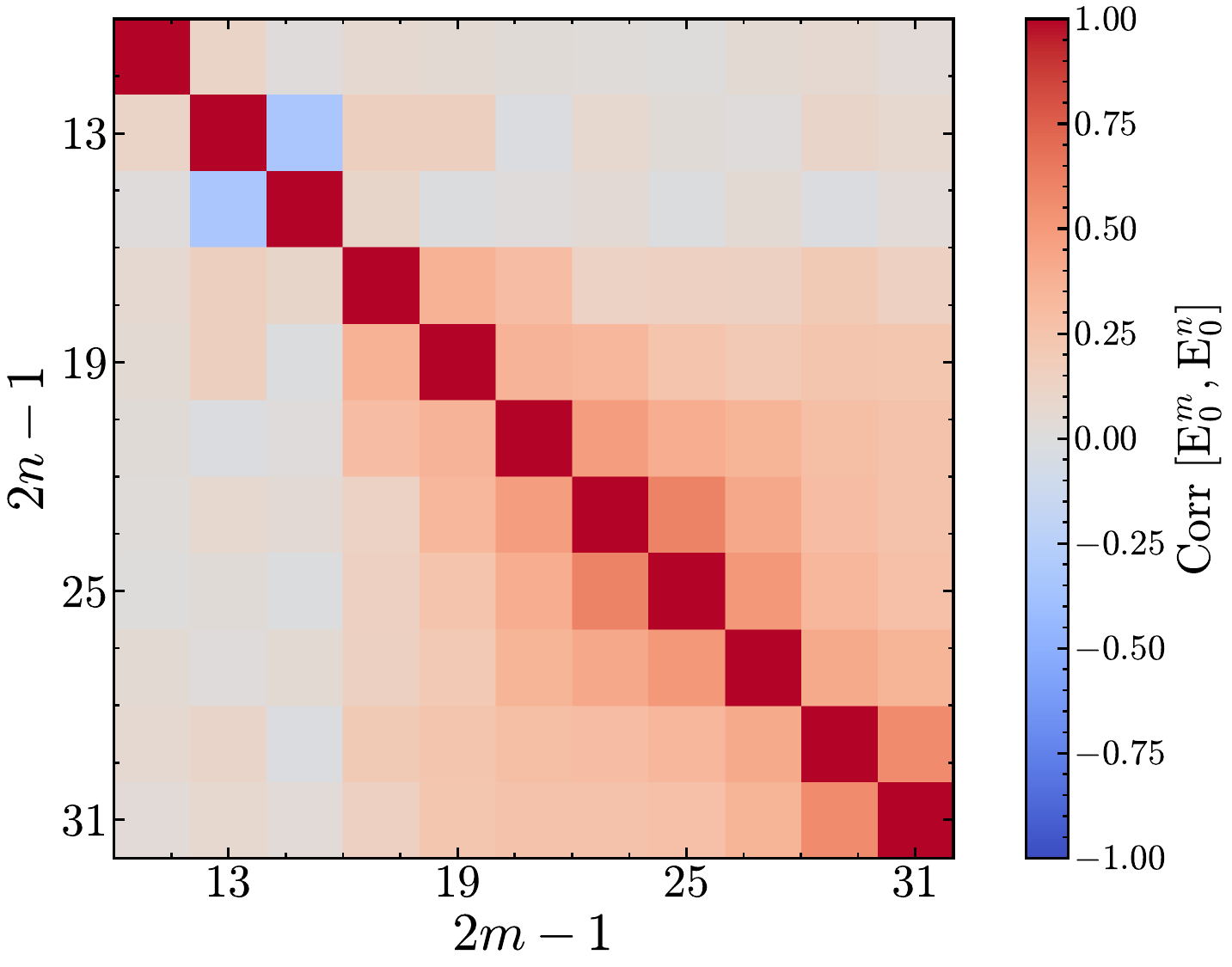}
    \caption{Correlation across different iterations of the TGEVP estimates for the ground-state energies of the deuteron, dineutron, $^3$He and $^4$He respectively from left to right. Results are shown for the SS setup with valence pion mass of $140$ MeV for two-nucleon systems and $300$ MeV for the helium nuclei.}
    \label{fig:TGEVP_corr}
\end{figure}

In Fig.~\ref{fig:TGEVP_corr}, we present the correlations among TGEVP estimates of the ground-state energies across different iterations for the deuteron, dineutron, $^3$He, and $^4$He (from left to right). The results are shown for the SS setup, with $m_{\pi} \approx 140~\mathrm{MeV}$ for the two-nucleon systems and $m_{\pi} \approx 300~\mathrm{MeV}$ for the helium nuclei. The strong cross-correlation observed among the TGEVP estimates at larger iterations indicates that the signal in the correlation functions has effectively saturated, reflecting the loss of independent information as excited-state contributions decay and statistical noise begins to dominate.

In Table~(\ref{Tab:Fit_results}), we list the extracted lowest-lying energy levels ($E_A$) and the corresponding energy shifts ($\Delta E_A$) for light nuclei up to ${}^4\mathrm{He}$. Results are shown for both symmetric (SS) and asymmetric (SP) source–sink setups, at three valence pion masses, and obtained using both the TGEVP method and multi-state fits.

\begin{table}[h!]
\centering
\begin{tabular}{cccccc}
\hline
\multirow{2}{*}{$m_{\pi}^{\text{val}}$(MeV)}& \multirow{2}{*}{} & \multirow{2}{*}{Nuclei} & \multicolumn{2}{c}{$E_A$ (MeV)} & \multirow{2}{*}{$-\Delta E_A$ (MeV)} \\
& & &  Mutli-State fit & TGEVP &  \\
\hline
\multirow{10}{*}{$695$}& \multirow{5}{*}{$\mathrm{SP}$} & $\mathrm{N}$ &$1475.90\pm1.17$ & $1477.25\pm1.79$ & $-$ \\
& & $\mathrm{NP}$ ($^3S_1$) & $2926.67\pm3.89$ & $2930.85\pm6.44$ & $24.58\pm6.78$\\
& & $\mathrm{NN}$ ($^1S_0$) & $2933.68\pm4.87$ & $2937.14\pm7.17$ &  $17.44\pm7.05$\\
& & $^3$He & $4342.75\pm11.94$& $4354.92\pm11.17$&   $77.25 \pm 11.15$\\
& & $^4$He & $5743.51\pm18.69$ & $5759.73\pm14.11$& $148.70\pm15.76$ \\
\cline{2-6}\cline{2-6}
& \multirow{5}{*}{$\mathrm{SS}$} & $\mathrm{N}$ &$1478.91\pm2.67$ & $1478.39\pm1.78$ & $-$ \\
& & $\mathrm{NP}$ ($^3S_1$) & $2949\pm3.37$& $2943.35\pm6.86$ & $13.49\pm7.31$\\
& & $\mathrm{NN}$ ($^1S_0$) & $2949.52\pm3.89$ & $2946.52\pm7.36$ & $10.35\pm7.88$\\
& & $^3$He & $4402.99\pm11.42$ & $4396.29\pm18.79$&  $39.04 \pm 19.33$\\
& & $^4$He & $5799.86\pm20.51$& $5818.66\pm24.11$& $96.57\pm22.56$\\
\hline\hline

\multirow{10}{*}{$295$}& \multirow{5}{*}{$\mathrm{SP}$} & $\mathrm{N}$ &  $1040.89\pm3.12$& $1042.62\pm4.37$  & $-$ \\
& & $\mathrm{NP}$ ($^3S_1$) & $2068.04\pm6.75$ & $2066.13\pm6.76$ & $17.65\pm8.75$\\
& & $\mathrm{NN}$ ($^1S_0$) & $2072.97\pm7.27$ & $2070.74\pm6.52$ & $13.28\pm8.81$\\
& & $^3$He & $3069.47\pm11.94$& $3076.13\pm10.39$&  $50.01 \pm 14.73$\\
& & $^4$He & $4070.39\pm19.21$ & $4073.67\pm18.47$& $94.16\pm24.07$ \\
\cline{2-6}\cline{2-6}
& \multirow{5}{*}{$\mathrm{SS}$} & $\mathrm{N}$ & $1045.05\pm3.89$ & $1045.53\pm3.33$ & $-$ \\
& & $\mathrm{NP}$ ($^3S_1$) &  $2069.59\pm7.01$ & $2069.23\pm7.74$ &  $19.03\pm8.97$\\
& & $\mathrm{NN}$ ($^1S_0$) &  $2075.83\pm7.45$ & $2074.74\pm8.95$ &  $14.42\pm9.79$\\
& & $^3$He & $3080.38\pm17.91$ & $3074.04\pm19.32$&   $56.46 \pm 19.26$\\
& & $^4$He & $4091.94\pm62.31$ & $4114.78\pm66.61$&  $62.83\pm67.16$\\
\hline\hline
\multirow{6}{*}{$140$}& \multirow{3}{*}{$\mathrm{SP}$} & $\mathrm{N}$ & $931.59\pm3.12$ & $931.93\pm4.71$ &  $-$ \\
& & $\mathrm{NP}$ ($^3S_1$) & $1858.51\pm14.80$ & $1860.44\pm18.31$ & $6.67\pm17.43$\\
& & $\mathrm{NN}$ ($^1S_0$) & $1892.52\pm23.11$ & $1890.08\pm22.25$ & $-23.48\pm21.48$\\
\cline{2-6}\cline{2-6}
& \multirow{3}{*}{$\mathrm{SS}$} & $\mathrm{N}$ &  $930.21\pm6.78$& $930.86\pm7.25$  & $-$ \\
& & $\mathrm{NP}$ ($^3S_1$) &  $1850.11\pm21.77$ & $1848.20\pm23.64$ &  $11.96\pm23.92$\\
& & $\mathrm{NN}$ ($^1S_0$) &  $1883.01\pm23.87$ & $1880.85\pm22.46$ &  $-18.18\pm25.06$\\
\hline\hline
\end{tabular}
\caption{Summary of extracted ground state energy levels ($E_A$) and energy shifts 
($\Delta E_A$) obtained using TGEVP estimates and multi-state fits for single nucleon and light nuclei at various valence pion masses and source-sink setups.}
\label{Tab:Fit_results}
\end{table}

\section{Comparison of smeared-smeared (SS) and smeared-point (SP) results\label{sec:SSvsSP}}

\begin{figure}[htpb!]
    \centering
    \includegraphics[scale=0.3]{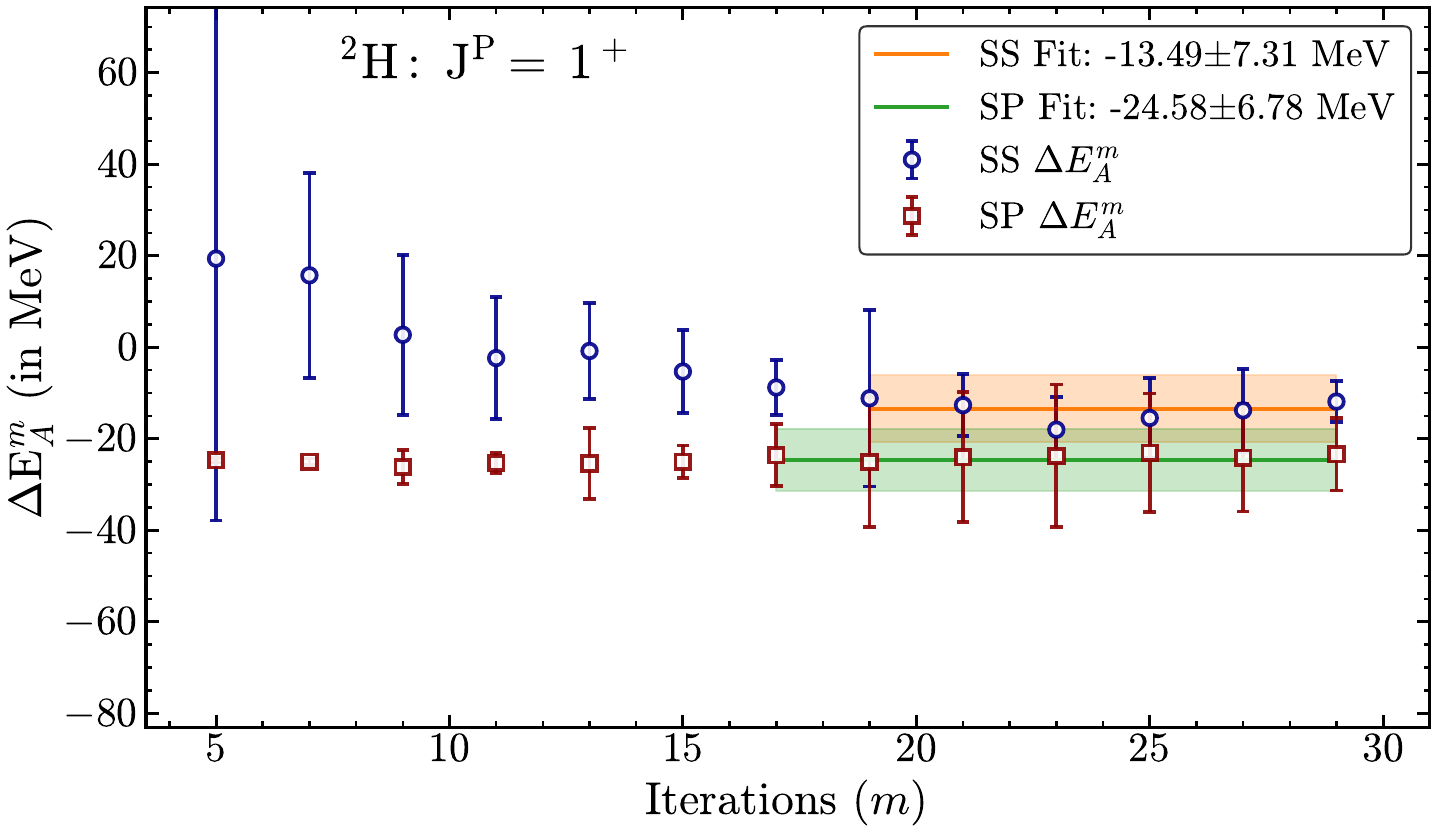}
    \includegraphics[scale=0.3]{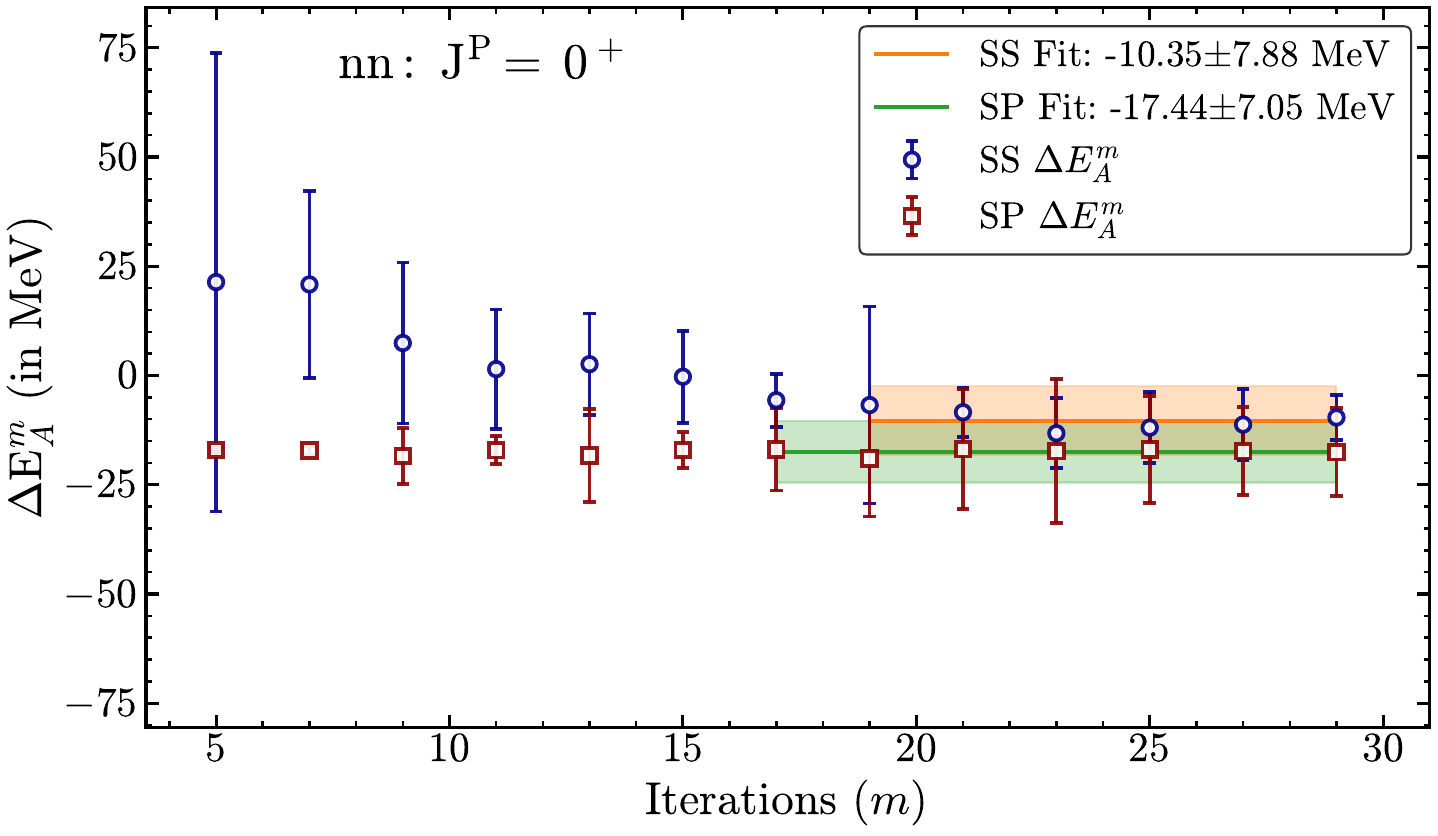}
    \includegraphics[scale=0.3]{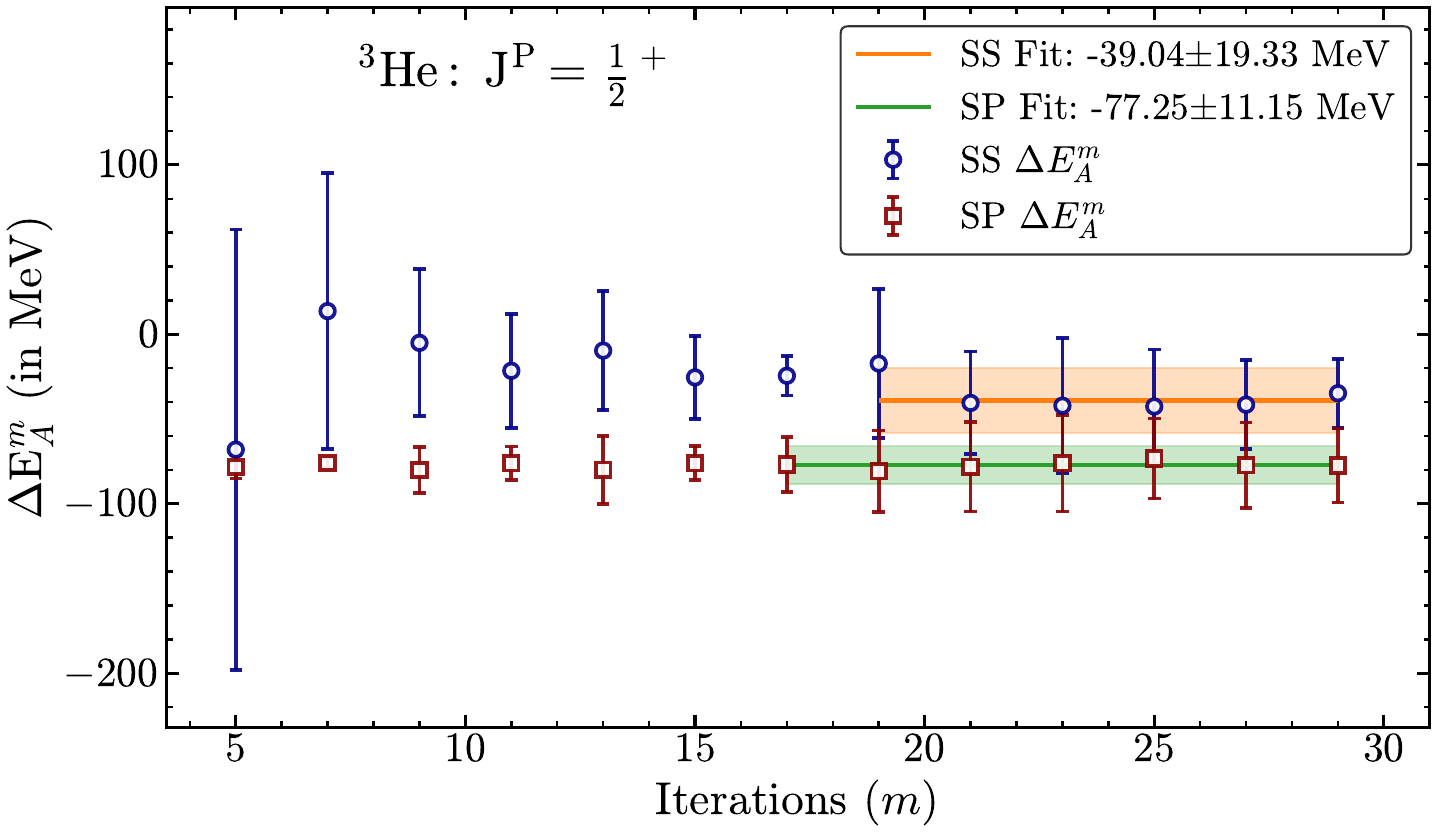}
    \includegraphics[scale=0.3]{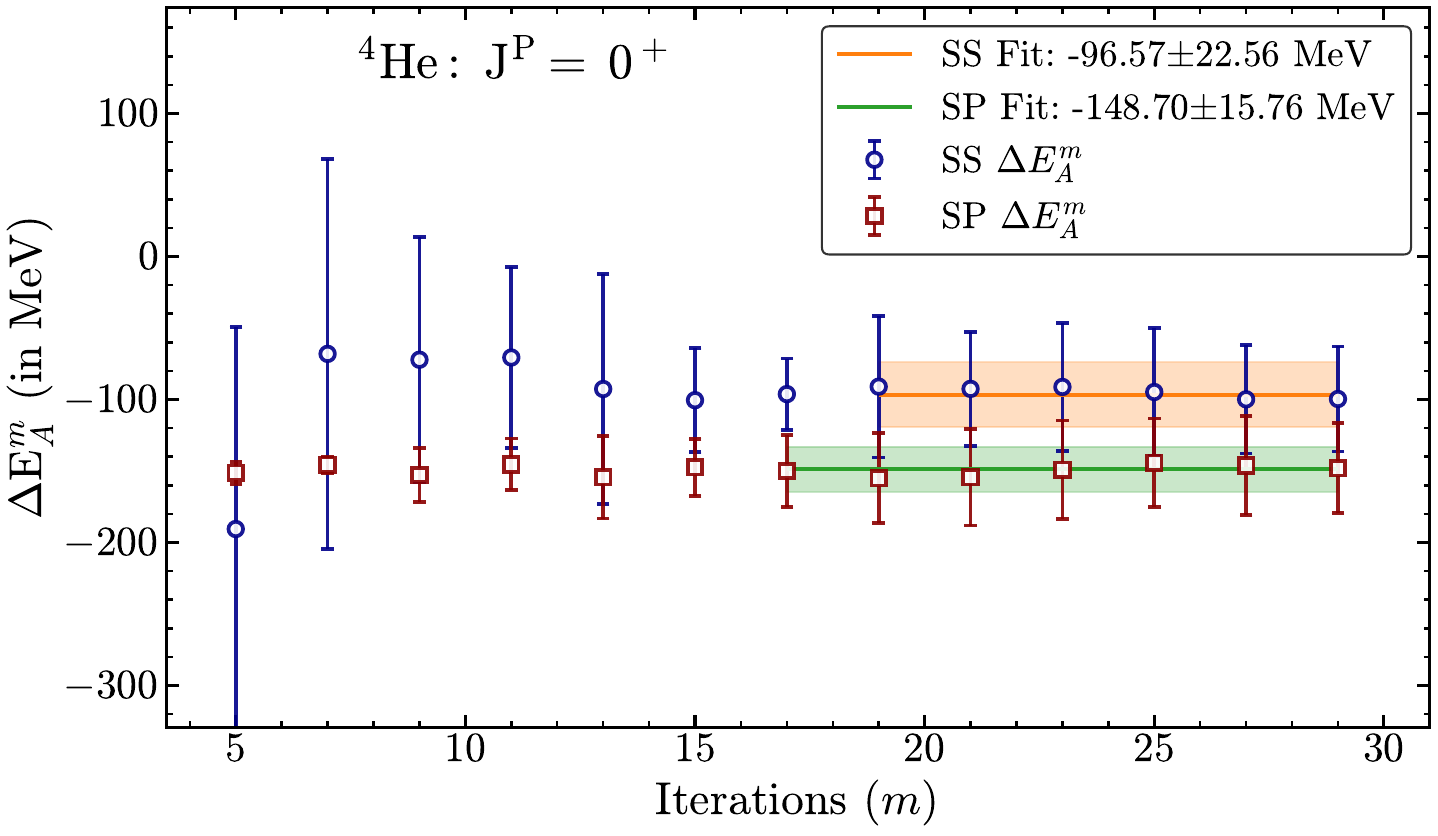}
    \caption{
The lowest-lying energy level $\Delta E_0^m$ from the TGEVP analysis of the ratio correlators is shown as a function of the iteration number $m$. Results are presented for the 
$^2$H (top-left), dineutron (top-right), $^3$He (bottom-left) and $^4$He (bottom-right) for valence pion mass $m_{\pi}\approx700$ MeV. The red points correspond to calculations with a smeared source and point sink, while the blue points correspond to a smeared source and smeared sink. The shaded bands indicate the fit results, with central values and $1\sigma$ uncertainties.
}
\label{fig:gevp_BE_mpi700}
\end{figure}

\begin{figure}[htpb!]
    \centering
    \includegraphics[scale=0.3]{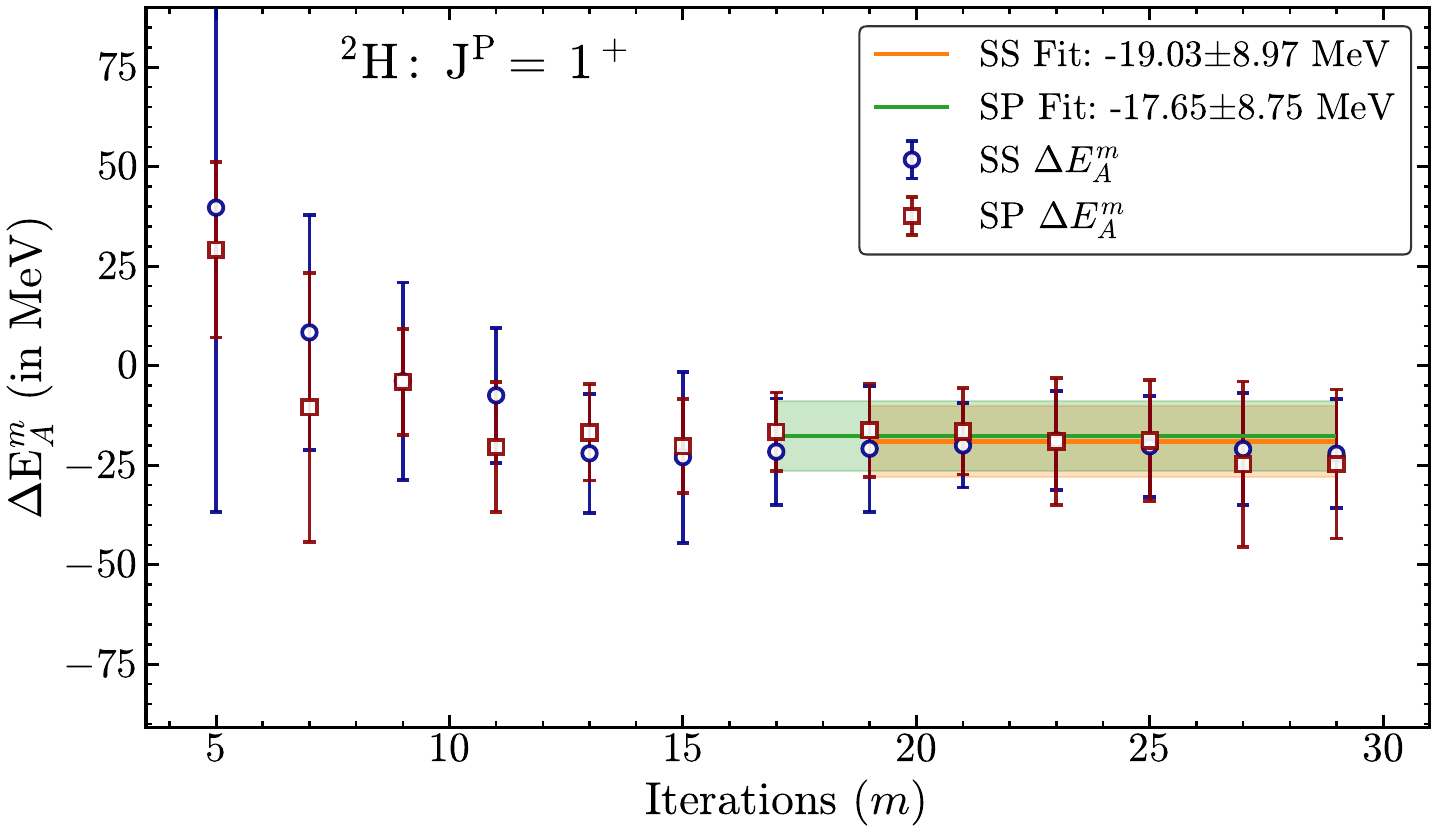}
    \includegraphics[scale=0.3]{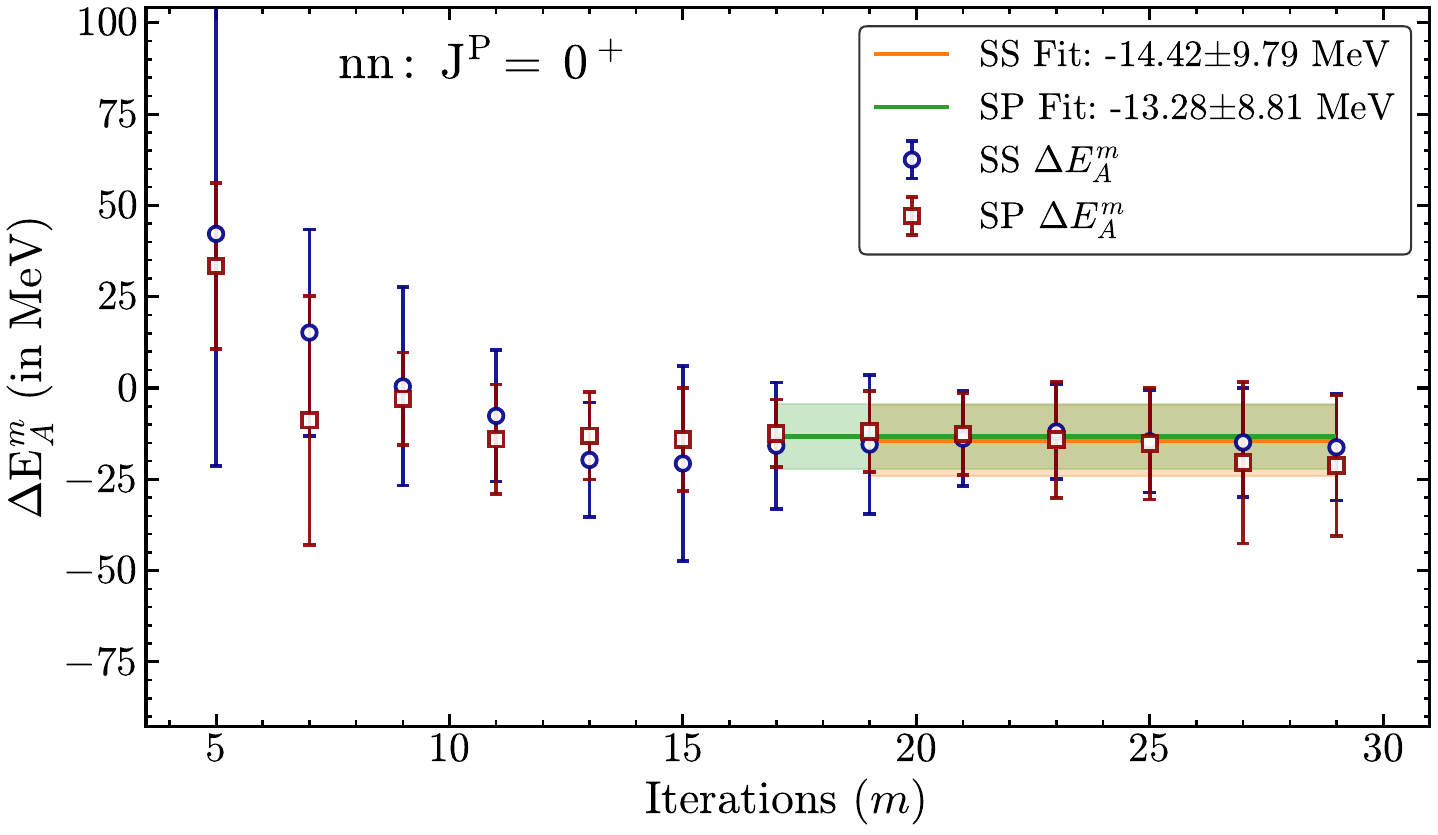}
    \includegraphics[scale=0.3]{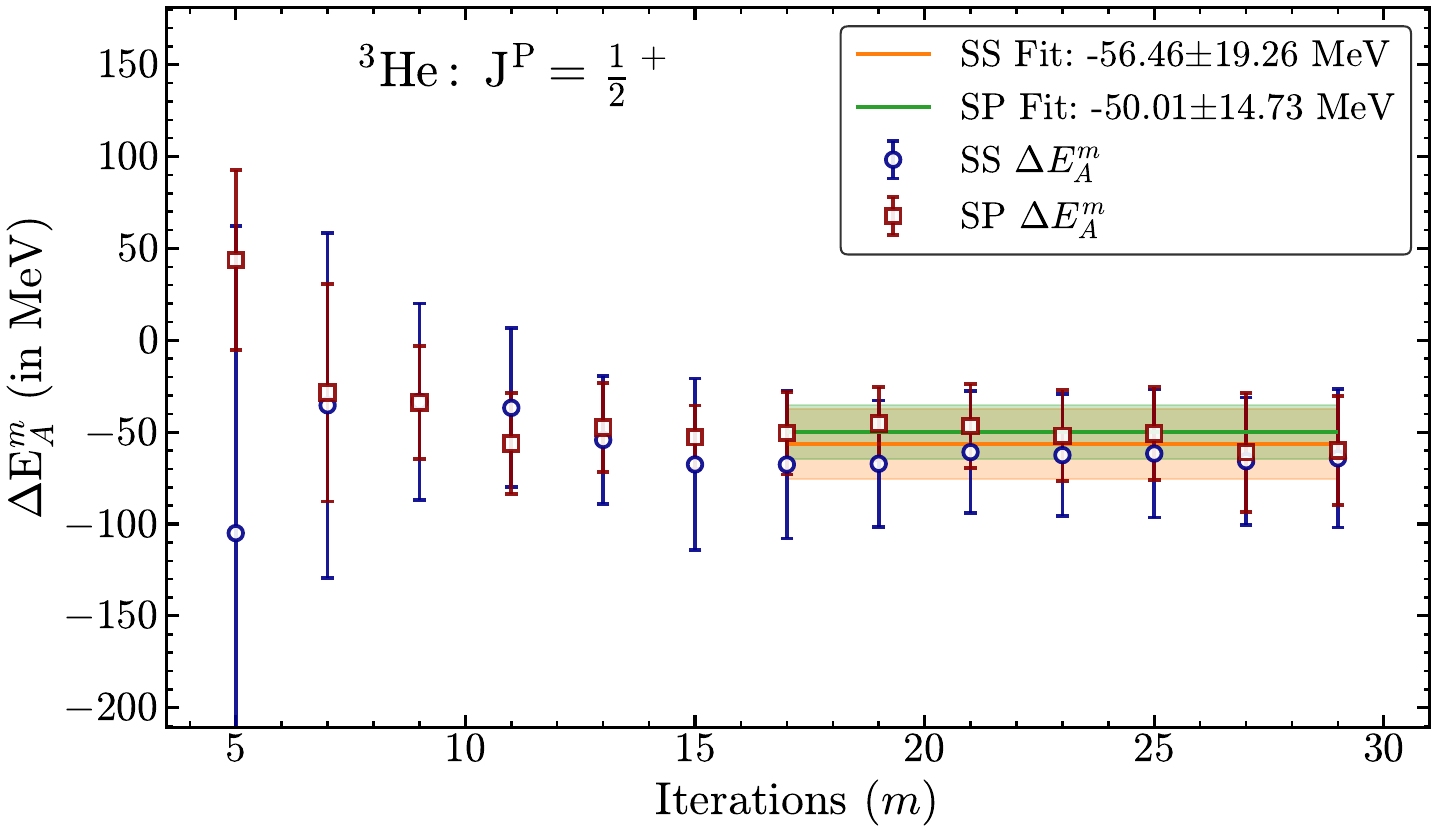}
    \includegraphics[scale=0.3]{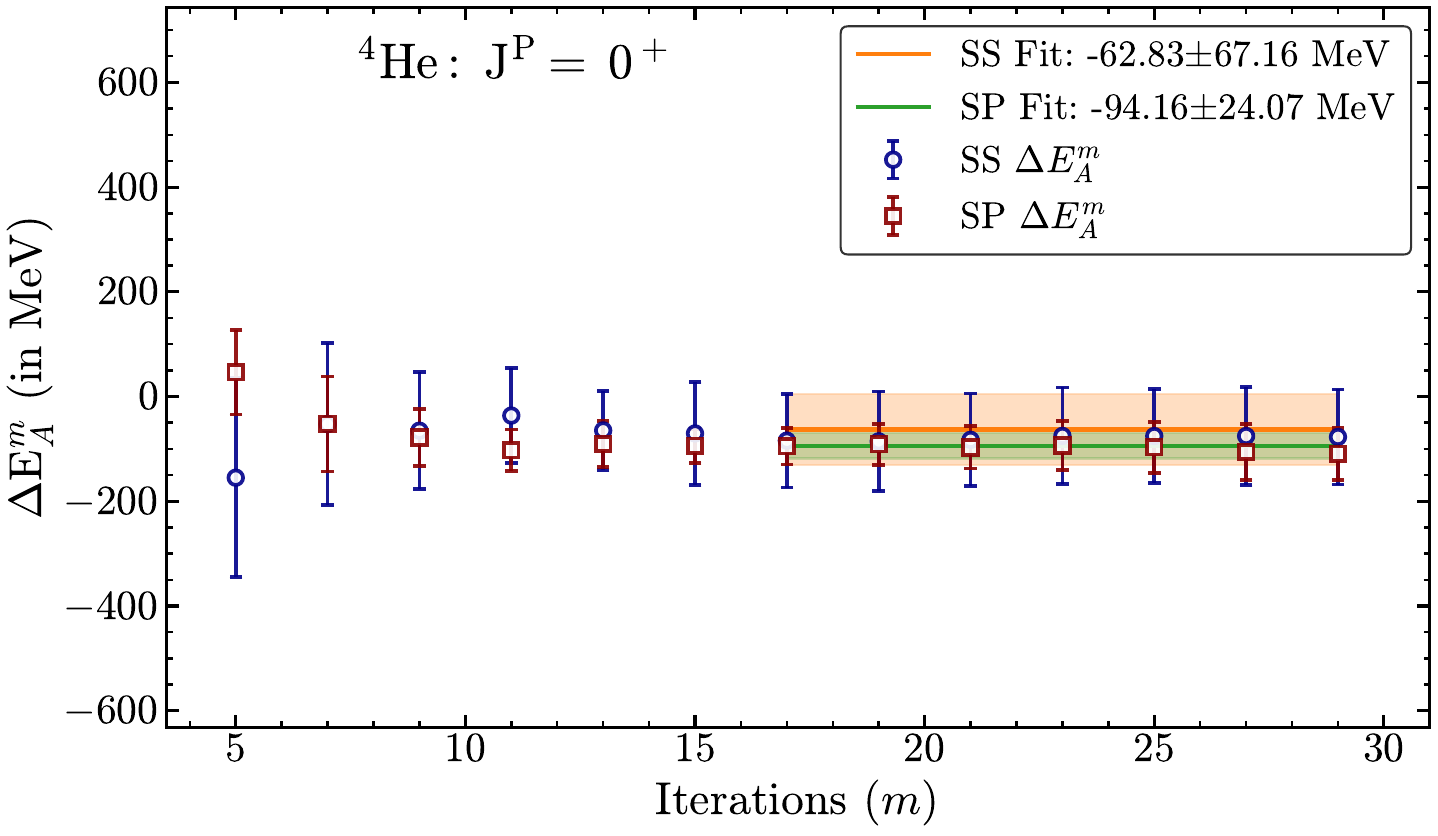}
    \caption{
Same as Fig. (\ref{fig:gevp_BE_mpi700}), however for valence pion mass $m_{\pi}\approx300$ MeV.
}
    \label{fig:gevp_BE_mpi300}
\end{figure}

\begin{figure}[htpb!]
    \centering
    \includegraphics[scale=0.35]{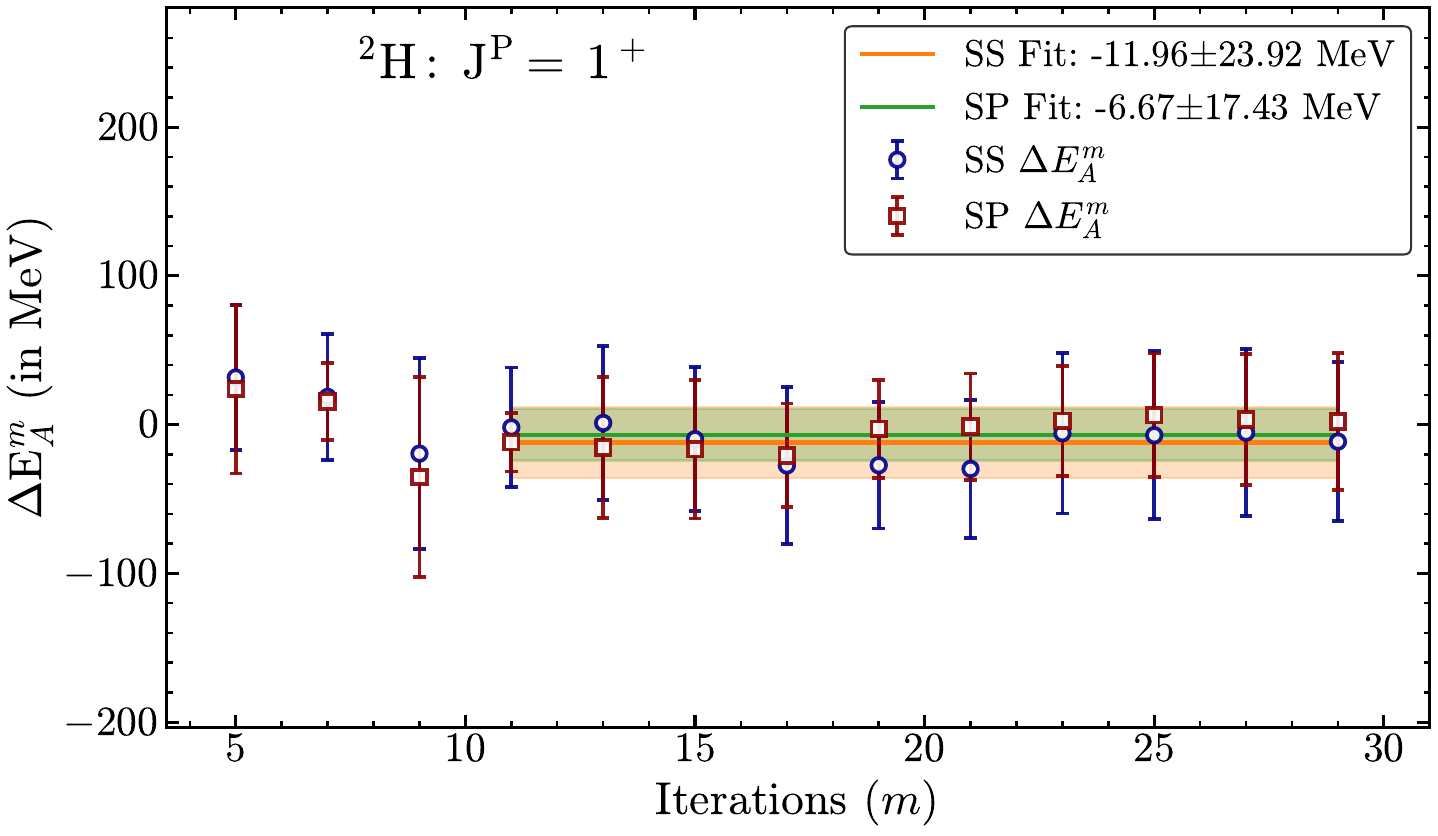}
    \includegraphics[scale=0.35]{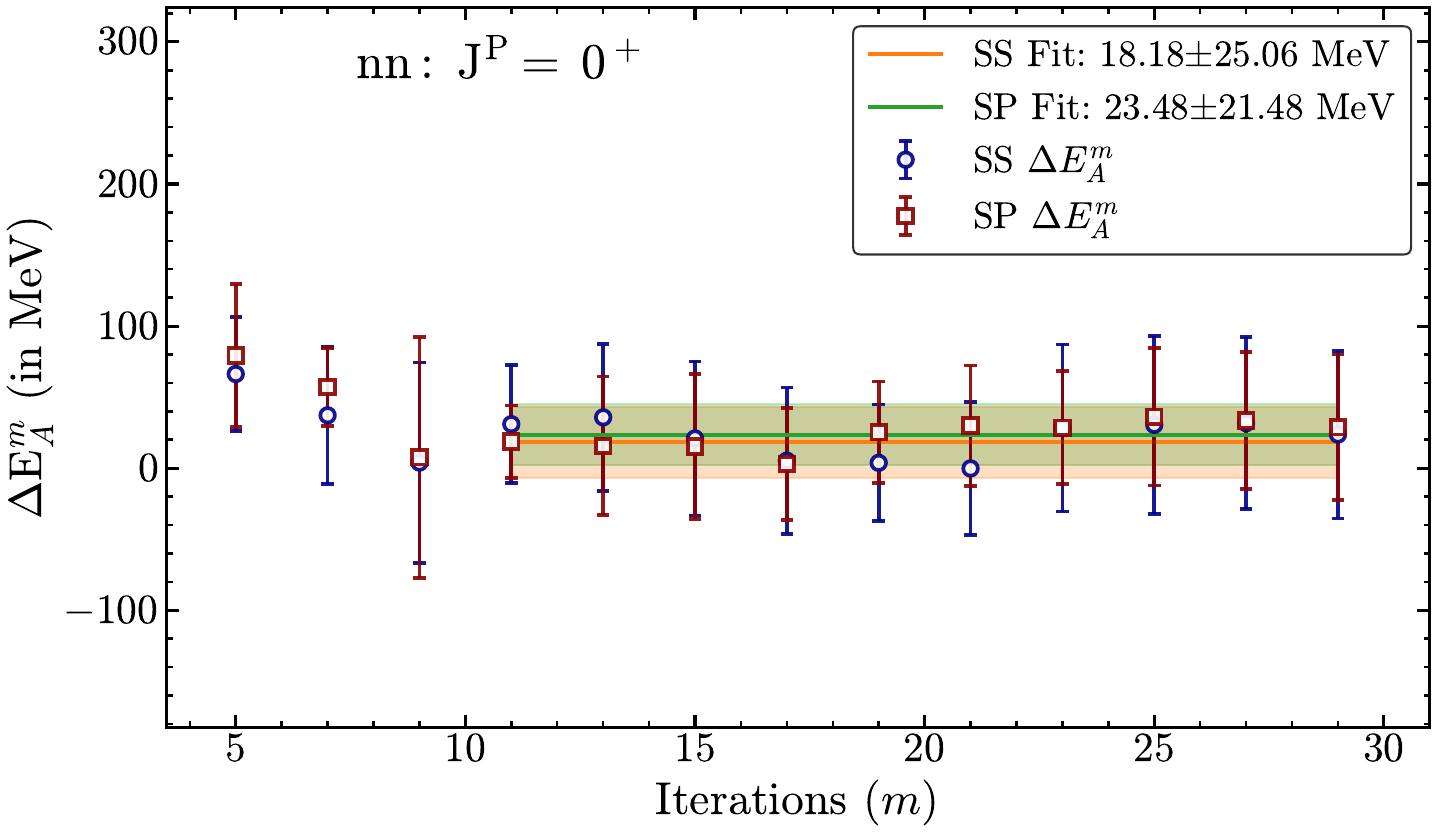}
    \caption{
Same as Fig. (\ref{fig:gevp_BE_mpi700}), however for valence pion mass $m_{\pi}\approx140$ MeV.
}
    \label{fig:gevp_BE_mpi140}
\end{figure}

In this section, we present a comparison of the finite-volume energy shifts,
$\Delta E_A := E_A - A \times m_N$, extracted for light nuclei up to
${}^4$He using smeared--smeared (SS) and smeared--point (SP) setups.
Here, $E_0$ denotes the lowest-lying energy level of the light nucleus,
$n$ is the mass number, and $m_N$ is the nucleon mass. In Figs.~(\ref{fig:gevp_BE_mpi700}, \ref{fig:gevp_BE_mpi300}, \ref{fig:gevp_BE_mpi140}), we present a
comparison of the extracted energy shifts $\Delta E_A$ for light nuclei
obtained using the TGEVP method in the SS and SP setups. The blue and brown
markers with error bars represent the extracted energy shifts from the SS and
SP setups, respectively, for different iterations of the TGEVP procedure.
The orange and light-green bands indicate the corresponding fit results to
the SS and SP TGEVP estimates, with the shaded regions representing the
$1\sigma$ uncertainties.

We find that the extracted lowest-lying energy levels from the SP and SS setups are consistent within uncertainties for lighter valence pion masses at $m_{\pi} \approx 300$ and $140$ MeV, across all light nuclei. However, at heavier valence pion masses, the SS setup yields significantly smaller energy shifts compared to the SP setup in all channels. This trend may arise from the fact that, with increasing pion mass, the spectrum of scattering states in a finite volume becomes progressively denser. Since the overlap factors can carry either positive or negative signs in an SP setup, the ground state becomes more susceptible to contamination from these nearby states. That may lead to an apparent plateau at short time separations, potentially resulting in negative energy shifts. Similar observations have been reported in \cite{BaSc2025, Detmold:2026vjx}.

\section{Nuclear Sigma Term}
\label{app:nuclear_sigma_term}

In this Appendix, we provide additional details on the extraction of the nuclear sigma-term ratio
\begin{equation}
    r_A \equiv \frac{\sigma_A}{\sigma_N}
    =
    \frac{\partial E_A/\partial m_q}{\partial m_N/\partial m_q}
    =
    \frac{\partial E_A}{\partial m_N},
    \label{eq:app_rA_def}
\end{equation}
where $\sigma_A$ and $\sigma_N$ denote the sigma terms of a nucleus and of the nucleon, respectively. Here $E_A$ is the ground-state energy extracted from the lattice correlation functions, and $m_N$ is the nucleon mass. The last equality follows from the Feynman--Hellmann theorem and allows us to determine $r_A$ directly from the quark-mass dependence of the spectrum.

In practice, we extract $r_A$ from linear fits of $E_A$ as a function of $m_N$. Two fit strategies are considered. The first uses only the lattice-QCD results. The second uses experimental nucleon and nuclear masses in place of the corresponding lattice values at the physical point. For ${}^3\mathrm{He}$ and ${}^4\mathrm{He}$, this supplements the two heavier-mass lattice points with an experimental anchor at the physical point, while for the deuteron the lattice value at the physical point is replaced by the corresponding experimental masses. Our central values and statistical uncertainties are taken from the fits using the experimental masses at the physical point, and the shift relative to the lattice-only fits is assigned as a systematic uncertainty. The quoted total uncertainty is obtained by combining the statistical and systematic contributions in quadrature.

\begin{figure}[h!]
    \centering
    \includegraphics[width=0.5\linewidth]{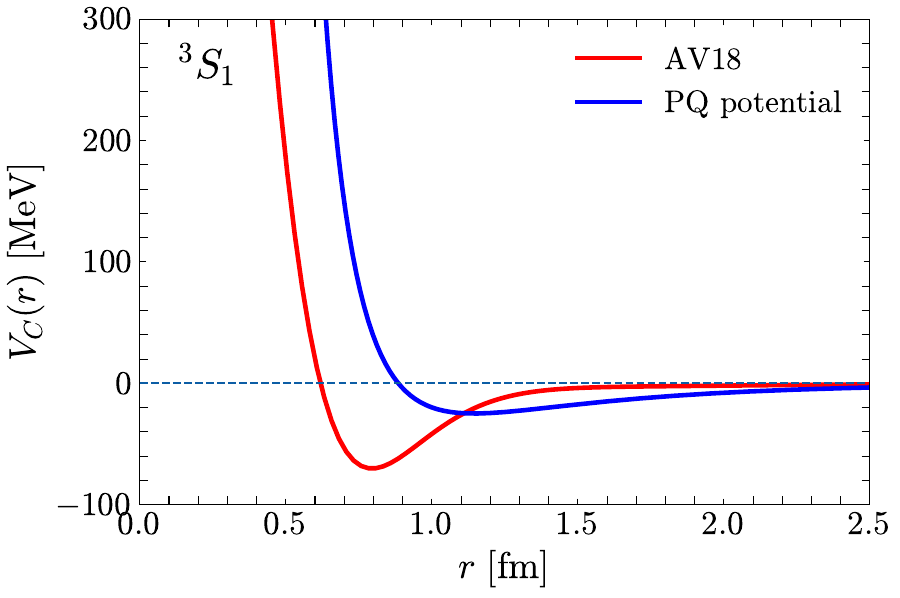}
    \caption{The Argonne V18 potential and partially quenched nucleon-nucleon potential plotted for $^3S_1$ channel at physical pion mass.}
    \label{fig:potentialplot}
\end{figure}

\begin{figure}[htb!]
    \centering
    
    % top row
    \includegraphics[width=0.32\linewidth]{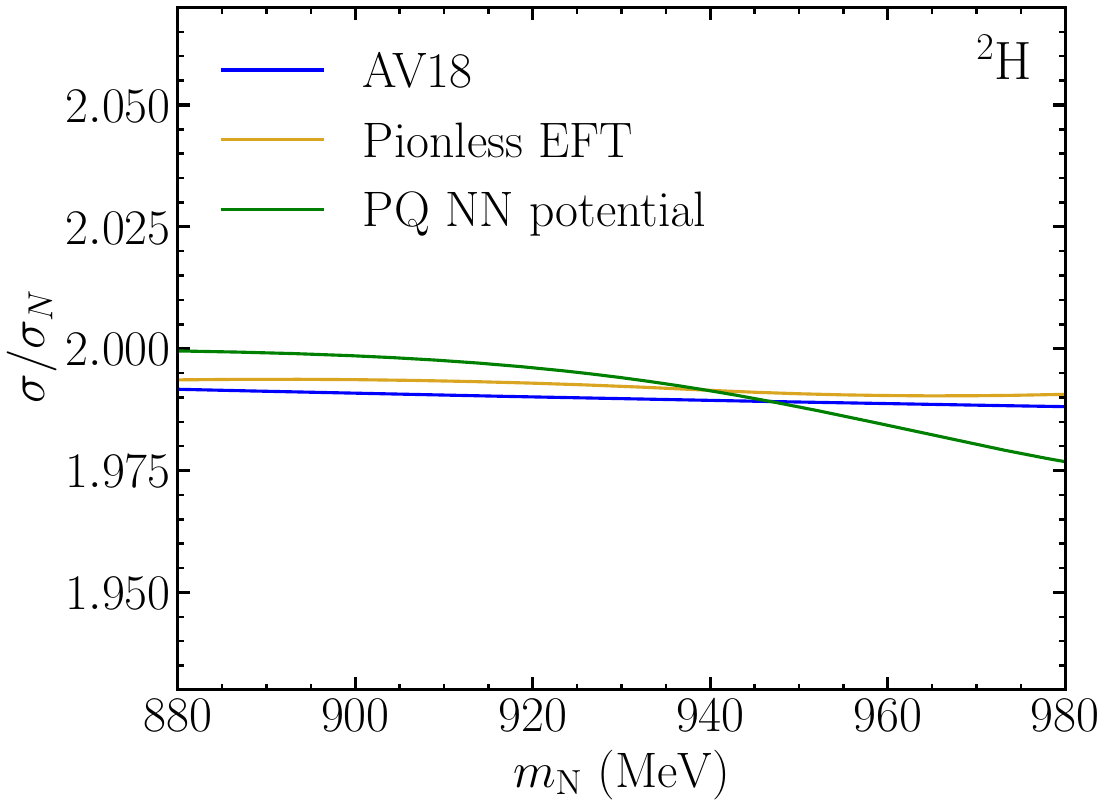}
    \includegraphics[width=0.32\linewidth]{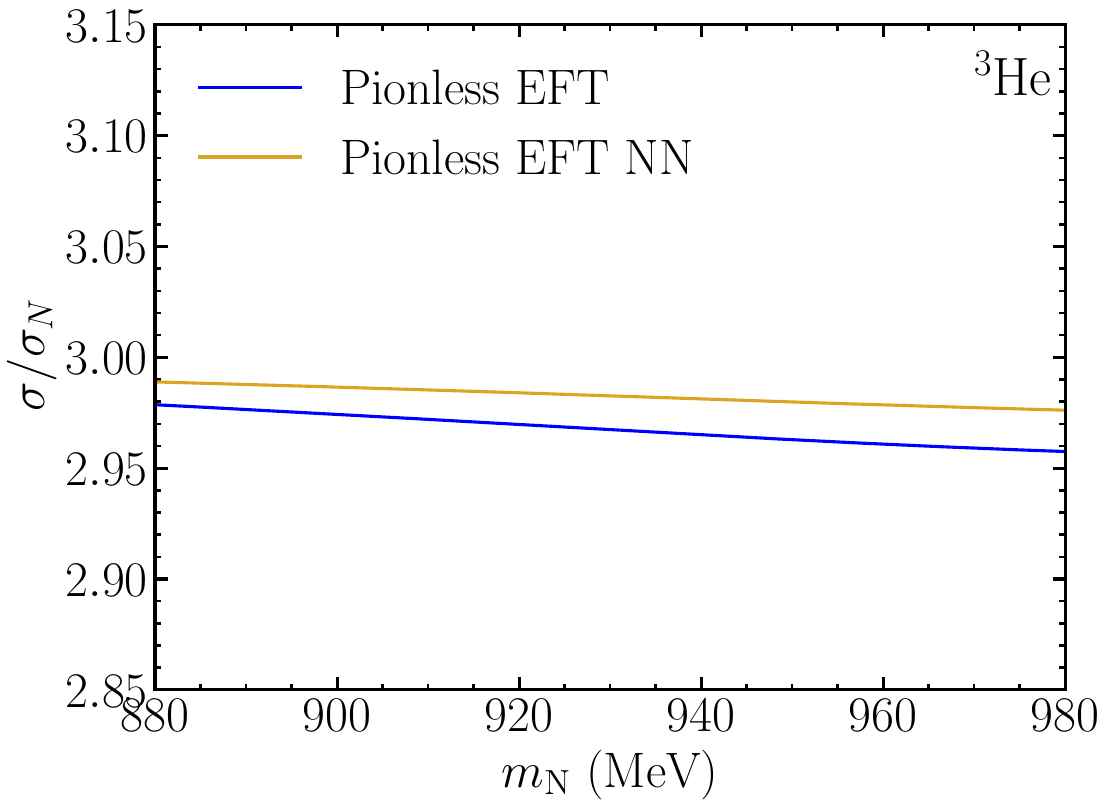}
    \includegraphics[width=0.32\linewidth]{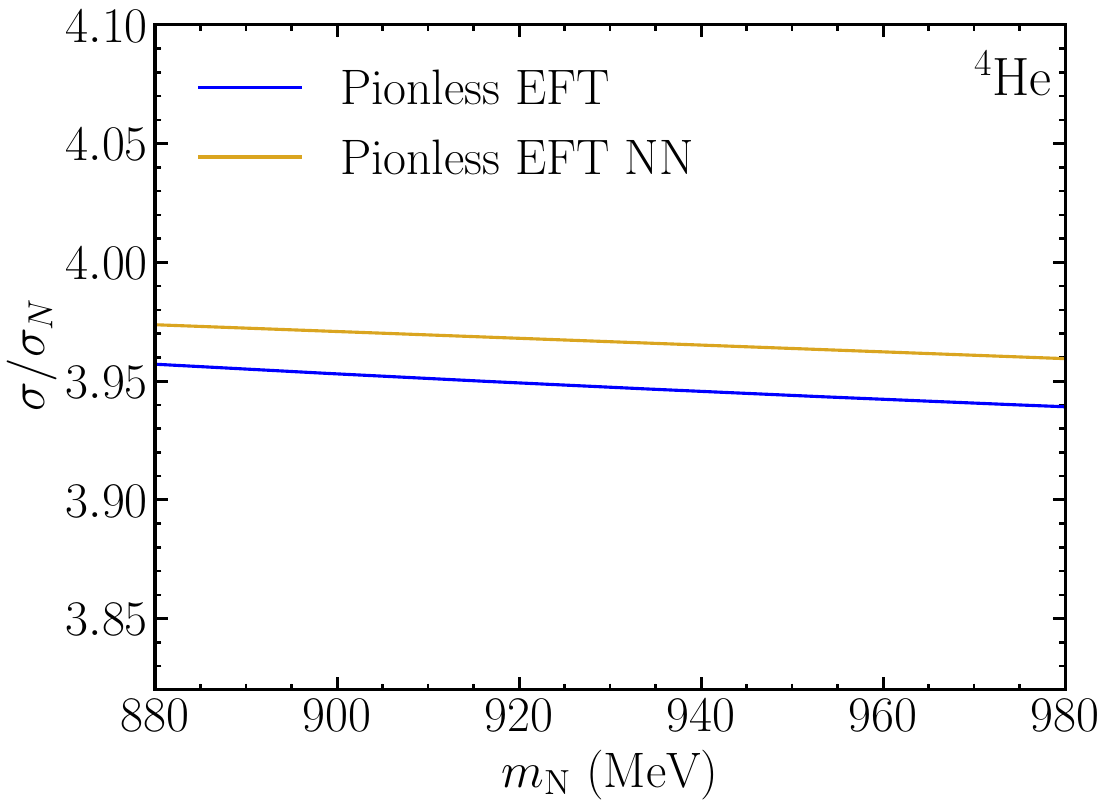}

    \caption{Sigma terms of light nuclei normalized to the corresponding single-nucleon value. The left panel shows results for the deuteron ($^2$H) obtained using the AV18 phenomenological potential, a partially quenched NN potential, and pionless EFT. The middle and right panels show the corresponding results for $^3$He and $^4$He, respectively, obtained within pionless EFT.}

    \label{fig:SummarySigmaFromPotential}
\end{figure}

To further benchmark our lattice-QCD determination, we compare the extracted sigma-term ratios with results obtained from phenomenological~\cite{AV18} and effective field theory approaches. In Fig.~\ref{fig:potentialplot}, we show the central part of the AV18 potential~\cite{AV18} and the partially quenched nucleon--nucleon potential~\cite{PQPotential,BEANE200391} in the $^3S_1$ channel at the physical pion mass. In Fig.~\ref{fig:SummarySigmaFromPotential}, we present $\sigma_A/\sigma_N$ for light nuclei normalized to the corresponding single-nucleon value. For the deuteron ($^2$H), results from the AV18 phenomenological potential, a partially quenched NN potential, and pionless EFT are shown, while for $^3$He and $^4$He we display the corresponding predictions from pionless EFT. This comparison provides a complementary perspective on the size of nuclear effects and the extent to which different approaches are consistent with the lattice-QCD extraction.

The resulting values of $\sigma_A/\sigma_N$ are collected in Table~\ref{tab:sigma_terms}, together with several benchmark expectations. Within uncertainties, the ratios remain close to $A$, indicating only modest deviations from naive additivity for the light nuclei considered here.

\begin{table}[h!]
    \centering
    \begin{tabular}{||c|c|c|c|c||}
         \hline
         \multirow{2}{*}{Light nuclei} & \multicolumn{4}{|c||}{$\sigma_A/\sigma_N$} \\\cline{2-5}
         & Lattice + experiment & Pionless EFT & Av18 & PQ EFT\\ \hline
         $^2\mathrm{H}$    & $1.975(6)(41)$ & $1.992$ & $1.990$ & $1.993$ \\
         $^3\mathrm{He}$   & $2.929(29)(126)$ & $2.966$ & $-$ & $-$ \\
         $^4\mathrm{He}$   & $3.876(34)(59)$ & $3.947$ & $-$ & $-$ \\
         \hline
    \end{tabular}
    \caption{Ratios of nuclear to nucleon sigma terms for the light nuclei considered in this work. The first uncertainty is statistical, and the second is the systematic uncertainty estimated from the difference between the fits with and without replacing the physical-point lattice values by experimental masses.}
    \label{tab:sigma_terms}
\end{table}

\begin{figure}[htb!]
    \centering
    \includegraphics[width=0.32\linewidth]{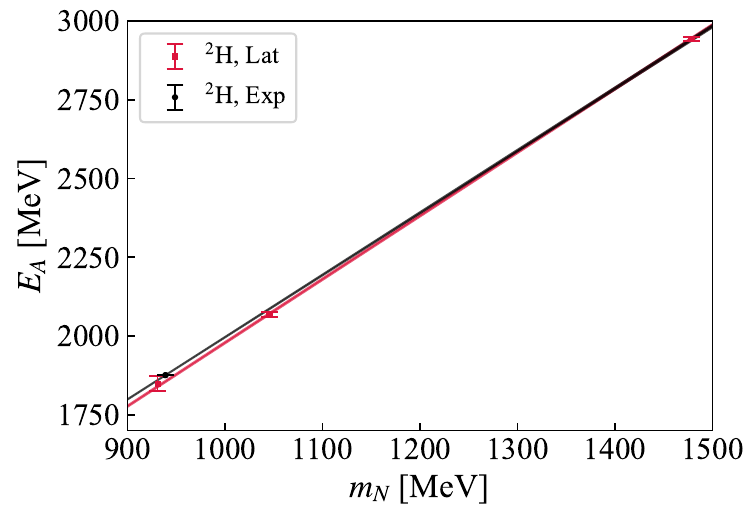}
    \includegraphics[width=0.32\linewidth]{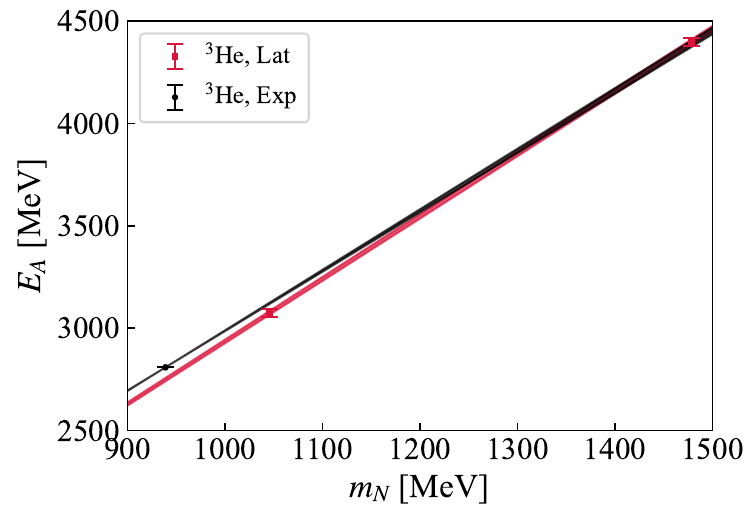}
    \includegraphics[width=0.32\linewidth]{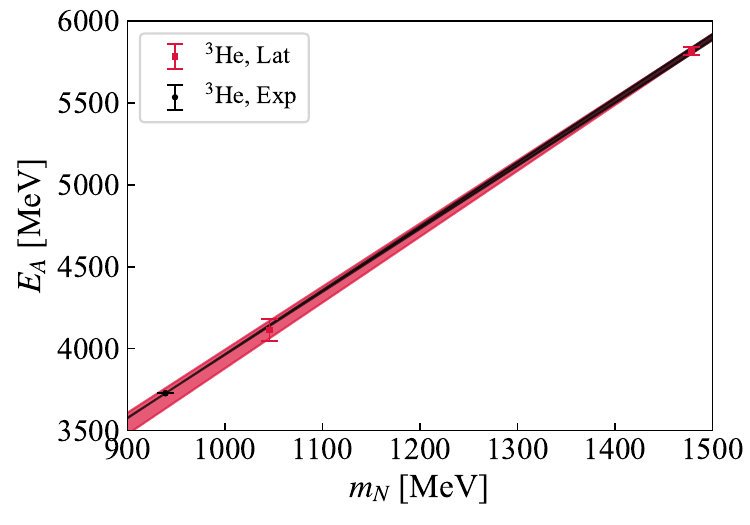}
    \caption{Linear fits used to extract $\sigma_A/\sigma_N$ from the dependence of $E_A$ on $m_N$ for the deuteron, ${}^3\mathrm{He}$, and ${}^4\mathrm{He}$. The red bands show fits using only lattice data, while the black bands show fits in which the physical-point lattice values are replaced by experimental masses.}
    \label{SigmaTermFit}
\end{figure}

Figure~\ref{SigmaTermFit} shows the corresponding linear fits of $E_A$ versus $m_N$ for the deuteron, ${}^3\mathrm{He}$, and ${}^4\mathrm{He}$. The red bands denote the fits using only lattice data, while the black bands denote the fits in which the physical-point lattice values are replaced by experimental masses. The comparison illustrates the stability of the extracted slope, which determines $\sigma_A/\sigma_N$, under this replacement.

\section{Nuclear trace anomaly}
\label{sec:nuclear_trace_anomaly}
In this Appendix, we provide additional details on the trace-anomaly decomposition of nuclear binding used in the main text. We work with the $\overline{\mathrm{MS}}$-renormalized QCD energy--momentum tensor (EMT) at scale $\mu$, whose trace is
\begin{equation}
T^{\nu}_{\nu} =
    \frac{\beta(\mu)}{2g(\mu)}\,(F^2)_R +
    \left[ 1 + \gamma_m(\mu) \right] (m\bar\psi\psi)_R \, ,
\label{eq:trace_anomaly}
\end{equation}
where $\beta(\mu)$ is the QCD beta function, $\gamma_m(\mu)$ is the quark-mass anomalous dimension, and $(F^2)_R$ and $(m\bar\psi\psi)_R$ denote the renormalized gluonic and quark-mass operators, respectively. Equation~\eqref{eq:trace_anomaly} is the standard form of the QCD trace anomaly in a mass-independent renormalization scheme~\cite{Nielsen:1977sy,Adler:1976zt,Collins:1976yq,Hatta:2018sqd}.

For numerical estimates, we use the perturbative expansions of $\beta(\mu)$ and $\gamma_m(\mu)$ through three loops~\cite{Hatta:2018sqd,Tanaka:2018nae},
\begin{equation}
\frac{\beta(g)}{2g}
= -\frac{\beta_0}{2}\frac{\alpha_s}{4\pi}
-\frac{\beta_1}{2}\left(\frac{\alpha_s}{4\pi}\right)^2
-\frac{\beta_2}{2}\left(\frac{\alpha_s}{4\pi}\right)^3 \,,
\qquad
\gamma_m
= \gamma_{m0}\frac{\alpha_s}{4\pi}
+\gamma_{m1}\left(\frac{\alpha_s}{4\pi}\right)^2
+\gamma_{m2}\left(\frac{\alpha_s}{4\pi}\right)^3 \,,
\label{eq:beta_gamma_expansion}
\end{equation}
with
\begin{align}
\beta_0 &= \frac{11}{3}C_A - \frac{2}{3}n_f \,,
\\
\beta_1 &= \frac{34}{3}C_A^{2} - 2C_F n_f - \frac{10}{3}C_A n_f \,,
\\
\beta_2 &= \frac{2857}{54}C_A^{3}
-\frac{1}{2}n_f\left(\frac{1415}{27}C_A^{2}+\frac{205}{9}C_A C_F-2C_F^{2}\right)
+\frac{1}{4}n_f^{2}\left(\frac{158}{27}C_A+\frac{44}{9}C_F\right)\,,
\\
\gamma_{m0} &= 6C_F \,,
\\
\gamma_{m1} &= 3C_F^{2} + \frac{97}{3}C_F C_A - \frac{10}{3}C_F n_f \,,
\\
\gamma_{m2} &= n_f\left[\left(-48\zeta(3)-\frac{556}{27}\right)C_A C_F
+\left(48\zeta(3)-46\right)C_F^{2}\right]
-\frac{129}{2}C_A C_F^{2}
+\frac{11413}{54}C_A^{2}C_F
-\frac{70}{27}C_F n_f^{2}
+129C_F^{3}\,.
\end{align}
Here $C_A=3$, $C_F=4/3$, and $\zeta(3)=1.2020569\ldots$.

\subsection{Trace-anomaly decomposition of nuclear binding}
The forward matrix element of the EMT in a spin-averaged hadron state satisfies
\begin{equation}
\langle P|T^{\mu\nu}|P\rangle = 2P^\mu P^\nu,
\qquad
\langle P|T^\nu_{\ \nu}|P\rangle = 2M^2,
\label{eq:massEMT}
\end{equation}
which yields the exact trace decomposition of the hadron mass. Using the standard definition
\begin{equation}
\langle A|(m\bar\psi\psi)_R|A\rangle \equiv 2M_A\,\sigma_A,
\label{eq:sigma_def_appendix}
\end{equation}
one obtains
\begin{equation}
E_A = E_A^{\sigma}(\mu) + E_A^{F^2}(\mu),
\label{eq:EA_trace_decomp}
\end{equation}
with
\begin{equation}
E_A^{\sigma}(\mu)=\bigl[1+\gamma_m(\mu)\bigr]\sigma_A,
\qquad
E_A^{F^2}(\mu)=\frac{\beta(\mu)}{2g(\mu)}\,
\frac{\langle A|F^2|A\rangle}{2M_A}.
\label{eq:EA_sigma_F2}
\end{equation}
Here $A=1$ corresponds to the nucleon, denoted by $N$, while $A>1$ denotes a nuclear bound state.

To expose the QCD origin of nuclear binding, we apply Eq.~\eqref{eq:EA_trace_decomp} to a nucleus and to the nucleon and subtract the two relations. Defining the total binding energy as
\begin{equation}
\Delta E_A \equiv E_A - A M_N,
\label{eq:binding_def_appendix}
\end{equation}
we obtain the trace-anomaly decomposition
\begin{equation}
\Delta E_A = \Delta E_A^{\sigma}(\mu) + \Delta E_A^{F^2}(\mu),
\label{eq:binding_decomp_appendix}
\end{equation}
where
\begin{equation}
\Delta E_A^{\sigma}(\mu)
=
\bigl[1+\gamma_m(\mu)\bigr]
\left(\sigma_A-A\sigma_N\right)
=
\bigl[1+\gamma_m(\mu)\bigr]\sigma_N
\left(\frac{\sigma_A}{\sigma_N}-A\right),
\label{eq:DeltaEA_sigma}
\end{equation}
and
\begin{equation}
\Delta E_A^{F^2}(\mu)
=
\frac{\beta(\mu)}{2g(\mu)}
\left[
\frac{\langle A|F^2|A\rangle}{2M_A}
-
A\,\frac{\langle N|F^2|N\rangle}{2M_N}
\right].
\label{eq:DeltaEA_F2}
\end{equation}

Equation~\eqref{eq:binding_decomp_appendix} separates nuclear binding into a contribution from explicit chiral symmetry breaking, $\Delta E_A^\sigma(\mu)$, and a gluonic contribution associated with the QCD trace anomaly, $\Delta E_A^{F^2}(\mu)$.

\begin{figure}[htb!]
  \centering
  \includegraphics[width=0.7\linewidth]{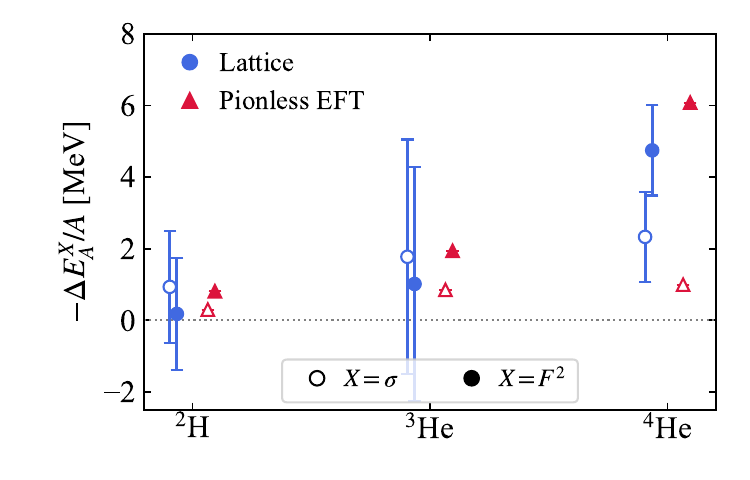}
  \caption{
  Decomposition of the binding energy per nucleon, $\Delta E/A$, at the $\overline{\mathrm{MS}}$ scale $\mu=2~\mathrm{GeV}$ into the quark-mass contribution, $\Delta E_{\sigma}$ (open symbols), and the gluonic trace-anomaly contribution, $\Delta E_{F^2}$ (filled symbols), for ${}^2\mathrm{H}$, ${}^3\mathrm{He}$, and ${}^4\mathrm{He}$. Circle symbols denote the lattice-QCD results, while triangle symbols denote the pionless-EFT benchmarks. In this figure, the decomposition is obtained using the phenomenological Roy--Steiner value of $\sigma_N$ from Ref.~\cite{Hoferichter:2023ptl}.
  }
  \label{fig:trace_decomp_ph}
\end{figure}

In the numerical analysis presented in the main text, we use our lattice determination of the ratio $\sigma_A/\sigma_N$, together with the lattice-QCD value of $\sigma_N$ from Ref.~\cite{Agadjanov:2023efe}, the physical nuclear masses, and the three-loop $\overline{\mathrm{MS}}$ expressions for $\gamma_m(\mu)$ and $\beta(\mu)$ given above. The same procedure is also applied to the pionless EFT values of $\sigma_A/\sigma_N$ shown in the figure for comparison. As an additional benchmark, we repeat the analysis using the phenomenological determination of $\sigma_N$ from the Roy--Steiner analysis of pion--nucleon scattering~\cite{Hoferichter:2023ptl}. The resulting decomposition at $\mu=2~\mathrm{GeV}$ is shown in Fig.~\ref{fig:trace_decomp_ph}.

\subsection{Trace-anomaly decomposition of nuclear mass\label{subsec:DetailsTADecomp}}

While the full EMT is renormalization-group invariant, its quark and gluon components are separately scale dependent because of operator mixing. At a fixed renormalization scale $\mu$, we decompose the trace of the EMT as
\begin{equation}
T^\nu_{\ \nu} = (T_q)^\nu_{\ \nu}(\mu) + (T_g)^\nu_{\ \nu}(\mu),
\label{eq:trace_split_qg}
\end{equation}
where the explicit $\mu$ dependence reflects the scheme dependence of the individual quark and gluon contributions, even though their sum reproduces the scale-independent trace of the full EMT.

Accordingly, the hadron mass can be written as
\begin{equation}
E_A = E_A^q(\mu) + E_A^g(\mu),
\label{eq:mass_decomp_qg}
\end{equation}
with
\begin{equation}
E_A^q(\mu) \equiv \frac{\langle A|(T_q)^\nu_{\ \nu}(\mu)|A\rangle}{2M_A},
\qquad
E_A^g(\mu) \equiv \frac{\langle A|(T_g)^\nu_{\ \nu}(\mu)|A\rangle}{2M_A}.
\label{eq:EA_qg_def}
\end{equation}

In the $\overline{\mathrm{MS}}$ scheme, the quark and gluon trace operators can be expressed perturbatively in terms of the renormalized operators $(F^2)_R$ and $(m\bar\psi\psi)_R$. Using the three-loop results of Ref.~\cite{Tanaka:2018nae}, we write
\begin{align}
(T_g)^\nu_{\ \nu}(\mu)
&=
\Bigl(
-0.437676\,\alpha_s(\mu)
-0.261512\,\alpha_s^2(\mu)
-0.183827\,\alpha_s^3(\mu)
\Bigr)\,(F^2)_R
\nonumber\\
&\quad+
\Bigl(
0.495149\,\alpha_s(\mu)
+0.776587\,\alpha_s^2(\mu)
+0.865492\,\alpha_s^3(\mu)
\Bigr)\,(m\bar\psi\psi)_R,
\label{eq:Tg_trace}
\\[4pt]
(T_q)^\nu_{\ \nu}(\mu)
&=
\Bigl(
0.0795775\,\alpha_s(\mu)
+0.0588695\,\alpha_s^2(\mu)
+0.0216037\,\alpha_s^3(\mu)
\Bigr)\,(F^2)_R
\nonumber\\
&\quad+
\Bigl(
1
+0.141471\,\alpha_s(\mu)
-0.00823495\,\alpha_s^2(\mu)
-0.0643511\,\alpha_s^3(\mu)
\Bigr)\,(m\bar\psi\psi)_R.
\label{eq:Tq_trace}
\end{align}
These expressions make explicit that, at finite $\mu$, the quark and gluon trace contributions each contain both gluonic and quark-mass operators.

Using the sigma term and gluonic matrix element determined in the previous subsection, we evaluate
\begin{equation}
E_A^q(\mu)=\frac{\langle A|(T_q)^\nu_{\ \nu}(\mu)|A\rangle}{2M_A},
\qquad
E_A^g(\mu)=\frac{\langle A|(T_g)^\nu_{\ \nu}(\mu)|A\rangle}{2M_A},
\end{equation}

\begin{figure}[htb!]
  \centering  \includegraphics[width=0.6\linewidth]{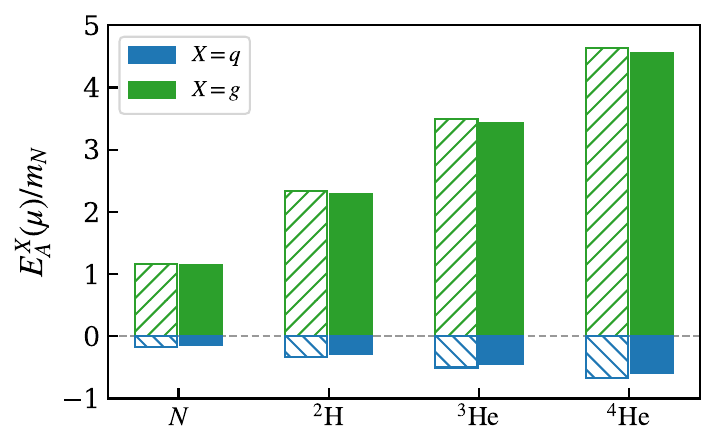}
  \caption{Trace-anomaly decomposition of the hadron mass at $\mu=2~\mathrm{GeV}$ into the quark contribution $E_A^q(\mu)$ and the gluon contribution $E_A^g(\mu)$ for ${}^2\mathrm{H}$, ${}^3\mathrm{He}$, and ${}^4\mathrm{He}$. Hatched bands are obtained using the lattice-QCD value of $\sigma_N$ from Ref.~\cite{Agadjanov:2023efe}, while filled bands use the Roy--Steiner value from Ref.~\cite{Hoferichter:2023ptl}.}
  \label{fig:trace_mass_decomp}
\end{figure}

The numerical results obtained using the lattice-QCD value of $\sigma_N$ from Ref.~\cite{Agadjanov:2023efe} (hatched bands) and the Roy--Steiner value from Ref.~\cite{Hoferichter:2023ptl} (filled bands) are shown in Fig.~\ref{fig:trace_mass_decomp}. A clear hierarchy is observed in all channels: the quark contribution $E_A^q(\mu)$ is negative and subleading, whereas the gluon contribution $E_A^g(\mu)$ is positive and provides the dominant share of the hadron mass in the trace decomposition. As $A$ increases from ${}^2\mathrm{H}$ to ${}^4\mathrm{He}$, both contributions vary smoothly, with the dominant growth carried by $E_A^g(\mu)$. The comparison between hatched and filled bands shows that this pattern is only mildly affected by the choice of $\sigma_N$ input.

%%%%%%%%%%%%%%%%%%%%%%%%%%%%%%%%%%%%%%%%%%%%%%%%%%%%%%%%%%%%%%%%%%%%%%%%%%%%%%%%%

%%%%%%%%%%%%%%%%%%%%%%%%%%%%%%%%%%%%%%%%%%%%%%%%%%%%%%%%%%

\clearpage
\bibliography{main_arXiv}

\end{document}